\documentclass[aps,12pt,final,notitlepage,oneside,onecolumn,nobibnotes,nofootinbib,superscriptaddress,noshowpacs,centertags]{revtex4}
\usepackage[T1]{fontenc}
\usepackage{amsthm}
\usepackage{amsmath}
\usepackage{graphicx}
\usepackage{amssymb}
\usepackage{esint}

\makeatletter
\@ifundefined{textcolor}{}
{%
 \definecolor{BLACK}{gray}{0}
 \definecolor{WHITE}{gray}{1}
 \definecolor{RED}{rgb}{1,0,0}
 \definecolor{GREEN}{rgb}{0,1,0}
 \definecolor{BLUE}{rgb}{0,0,1}
 \definecolor{CYAN}{cmyk}{1,0,0,0}
 \definecolor{MAGENTA}{cmyk}{0,1,0,0}
 \definecolor{YELLOW}{cmyk}{0,0,1,0}
 }
\theoremstyle{plain}
\theoremstyle{plain}
\newtheorem{thm}{Theorem}
  \theoremstyle{definition}
  \newtheorem{defn}[thm]{Definition}
  \theoremstyle{remark}
  \newtheorem{rem}[thm]{Remark}
  \theoremstyle{definition}
  \newtheorem*{example*}{Example}
 \theoremstyle{definition}
  \newtheorem{example}[thm]{Example}
  \theoremstyle{plain}
  \newtheorem{cor}[thm]{Corollary}
  \theoremstyle{plain}
  \newtheorem{lem}[thm]{Lemma}

\usepackage[all]{xy} 
\usepackage{amscd}
\usepackage{wasysym}
\makeatletter 
\newcommand{\xyR}[1]{ \makeatletter
\xydef@\xymatrixrowsep@{#1} \makeatother} 
\newcommand{\xyC}[1]{ \makeatletter
\xydef@\xymatrixcolsep@{#1} \makeatother} 
\entrymodifiers={++[ ][F]} 

\newcommand{\h}[1]{\hspace{#1 truein}}  
\newcommand{\sss}{\scriptscriptstyle} 

\newcommand{\freccia}{\longrightarrow} 
\newcommand{\xfrecciad}[1]{\xrightarrow{\displaystyle{\ \ \ #1\ \ \ }}}

\newcommand{\eps}{\varepsilon} 
\renewcommand{\phi}{\varphi} 
\newcommand{\field}[1]{\mathbb{#1}}
\newcommand{\R}{\field{R}}                        
\newcommand{\ER}{{^\bullet\R}}                    
\newcommand{\N}{\field{N}}                        
\newcommand{\diff}[1]{\,{\rm d}#1}

\newcommand{\st}[1]{{^\circ #1}} 
\newcommand{\Nil}{\mathcal{N}} 
\newcommand{\Cc}{\mathcal{C}} 
\newcommand{\D}{\mathcal{D}} 
\newcommand{\ext}[1]{{}^\bullet #1} 

\newcommand{\qedWithFinalEq}{\phantom{|}\hfill \vrule height 1.5ex width 1.3ex depth-.2ex}
\newcommand{\qedNoNewLine}{\\\phantom{|}\hfill \vrule height 1.5ex width 1.3ex depth-.2ex}
\newcommand{\ptind}{\displaystyle \mathop {\ldots\ldots\,}} 
\newcommand{\pti}{:\ } 
\newcommand{\then}{\quad \Longrightarrow \quad}
\newcommand{\DIff}{ \quad\ :\h{-0.02}\iff \quad } 

\makeatother

\begin{document}

\title{Infinitesimals without Logic}

\author{Paolo Giordano}

\affiliation{Universit\`a della Svizzera italiana}

\email{paolo.giordano@usi.ch}
\begin{abstract}
We introduce the ring of Fermat reals, an extension of the real field
containing nilpotent infinitesimals. The construction takes inspiration
from Smooth Infinitesimal Analysis (SIA), but provides a powerful
theory of actual infinitesimals without any need of a background in
mathematical logic. In particular, on the contrary with respect to
SIA, which admits models only in intuitionistic logic, the theory
of Fermat reals is consistent with classical logic. We face the problem
to decide if the product of powers of nilpotent infinitesimals is
zero or not, the identity principle for polynomials, the definition
and properties of the total order relation. The construction is highly
constructive, and every Fermat real admits a clear and order preserving
geometrical representation. Using nilpotent infinitesimals, every
smooth functions becomes a polynomial because in Taylor's formulas
the rest is now zero. Finally, we present several applications to
informal classical calculations used in Physics: now all these calculations
become rigorous and, at the same time, formally equal to the informal
ones. In particular, an interesting rigorous deduction of the wave
equation is given, that clarifies how to formalize the approximations
tied with Hook's law using this language of nilpotent infinitesimals.
\end{abstract}
\maketitle
\tableofcontents{}

\section{\label{sec:IntroductionAndGeneralProblem}Introduction and general
problem}

Frequently in work by physicists it is possible to find informal calculations
like \begin{equation}
\frac{1}{\sqrt{1-{\displaystyle \frac{v^{2}}{c^{2}}}}}=1+\frac{v^{2}}{2c^{2}}\qquad\qquad\sqrt{1-h_{44}(x)}=1-\frac{1}{2}h_{44}(x)\label{eq:EinsteinInfinitesimal}\end{equation}
 with explicit use of infinitesimals $v/c\ll1$ or $h_{44}(x)\ll1$
such that e.g. $h_{44}(x)^{2}=0$. For example \citet{Ein} wrote
the formula (using the equality sign and not the approximate equality
sign $\simeq$) \begin{equation}
f(x,t+\tau)=f(x,t)+\tau\cdot\frac{\partial f}{\partial t}(x,t)\label{eq:EinsteinDerivationFormula}\end{equation}
 justifying it with the words {}``\emph{since $\tau$ is very small}'';
the formulas \eqref{eq:EinsteinInfinitesimal} are a particular case
of the general \eqref{eq:EinsteinDerivationFormula}. \citet{Dir}
wrote an analogous equality studying the Newtonian approximation in
general relativity.

Using this type of infinitesimals we can write an \emph{equality},
in some infinitesimal neighborhood, between a smooth function and
its tangent straight line, or, in other words, a Taylor's formula
without remainder.

There are obviously many possibilities to formalize this kind of intuitive
reasonings, obtaining a more or less good dialectic between informal
and formal thinking, and indeed there are several theories of actual
infinitesimals (from now on, for simplicity, we will say {}``infinitesimals''
instead of {}``actual infinitesimals'' as opposed to {}``potential
infinitesimals''). Starting from these theories we can see that we
can distinguish between two type of definitions of infinitesimals:
in the first one we have at least a ring $R$ containing the real
field $\R$ and infinitesimals are elements $\eps\in R$ such that
$-r<\eps<r$ for every positive standard real $r\in\R_{>0}$. The
second type of infinitesimal is defined using some algebraic property
of nilpotency, i.e. $\eps^{n}=0$ for some natural number $n\in\N$.
For some ring $R$ these definitions can coincide, but anyway they
lead, of course, only to the trivial infinitesimal $\eps=0$ if $R=\R$.

However these definitions of infinitesimals correspond to theories
which are completely different in nature and underlying ideas. Indeed
these theories can be seen in a more interesting way to belong to
two different classes. In the first one we can put theories that need
a certain amount of non trivial results of mathematical logic, whereas
in the second one we have attempts to define sufficiently strong theories
of infinitesimals without the use of non trivial results of mathematical
logic. In the first class we have Non-Standard Analysis (NSA) and
Synthetic Differential Geometry (SDG, also called Smooth Infinitesimal
Analysis, see e.g. \citet{Bel,Koc,Lav,Mo-Re}), in the second one
we have, e.g., Weil functors (see \citet{Kr-Mi2}), Levi-Civita fields
(see \citet{Sha,Berz2}), surreal numbers (see \citet{Con2,Ehr}),
geometries over rings containing infinitesimals (see \citet{Ber}).
More precisely we can say that to work in NSA and SDG one needs a
formal control deeply stronger than the one used in {}``standard
mathematics''. Indeed to use NSA one has to be able to formally write
the sentences one needs to use the transfer theorem. Whereas SDG does
not admit models in classical logic, but in intuitionistic logic only,
and hence we have to be sure that in our proofs there is no use of
the law of the excluded middle, or e.g. of the classical part of De
Morgan's law or of some form of the axiom of choice or of the implication
of double negation toward affirmation and any other logical principle
which is not valid in intuitionistic logic. Physicists, engineers,
but also the greatest part of mathematicians are not used to have
this strong formal control in their work, and it is for this reason
that there are attempts to present both NSA and SDG reducing as much
as possible the necessary formal control, even if at some level this
is technically impossible (see e.g. \citet{Hen}, and \citet{Be-DN1,Be-DN2}
for NSA; \citet{Bel} and \citet{Lav} for SDG, where using an axiomatic
approach the authors try to postpone the very difficult construction
of an intuitionistic model of a whole set theory using Topos).

On the other hand NSA is essentially the only theory of infinitesimals
with a discrete diffusion and a sufficiently great community of working
mathematicians and published results in several areas of mathematics
and its applications, see e.g. \citet{Al-Fe-Ho-Li}. SDG is the only
theory of infinitesimals with non trivial, new and published results
in differential geometry concerning infinite dimensional spaces like
the space of all the diffeomorphisms of a generic (e.g. non compact)
smooth manifold. In NSA we have only few results concerning differential
geometry. Other theories of infinitesimals have not, at least up to
now, the same formal strength of NSA or SDG or the same potentiality
to be applied in several different areas of mathematics.

Our main aim, of which the present work represents a first step, is
to find a theory of infinitesimals within {}``standard mathematics''
(in the precise sense explained above of a formal control more {}``standard''
and not so strong as the one needed e.g. in NSA or SDG) with results
comparable with those of SDG, without forcing the reader to learn
a strong formal control of the mathematics he/she is doing. Because
it has to be considered inside {}``standard mathematics'', our theory
of infinitesimals must be compatible with classical logic.

Concretely, the idea of the present work is to by-pass the impossibility
theorem about the incompatibility of SDG with classical logic that
forces SDG to find models within intuitionistic logic.

Another point of view about present theories of infinitesimals is
that, in spite of the fact that frequently they are presented using
opposed motivations, they lacks the intuitive interpretation of what
the powerful formalism permits to do. For some concrete example in
this direction, see \citet{Gio4}. Another aim of the present work
is to construct a new theory of infinitesimals preserving always a
very good dialectic between formal properties and intuitive interpretation.

More technically we want to show that it is possible to extend the
real field adding nilpotent infinitesimals, arriving at an enlarged
real line $\ER$, by means of a very simple construction completely
inside {}``standard mathematics''. Indeed to define the extension
$\ER\supset\R$ we shall use elementary analysis only. To avoid misunderstandings
is it important to clarify that the purpose of the present work is
not to give an alternative foundation of differential and integral
calculus (like NSA), but to obtain a theory of nilpotent infinitesimals
as a first step for the foundation of a smooth ($\Cc^{\infty}$) differential
geometry. For some preliminary results in this direction, see \citet{Gio4}.

\section{Motivations for the name {}``Fermat reals''}

It is well known that historically two possible reductionist constructions
of the real field starting from the rationals have been made. The
first one is Dedekind's order completion using sections of rationals,
the second one is Cauchy's metric space completion. Of course there
are no historical reason to attribute our extension $\ER\supset\R$
of the real field, to be described below, to Fermat, but there are
strong motivations to say that, probably, he would have liked the
underlying spirit and some properties of our theory. For example: 
\begin{enumerate}
\item a formalization of Fermat's infinitesimal method to derive functions
is provable in our theory. We recall that Fermat's idea was, roughly
speaking and not on the basis of an accurate historical analysis which
goes beyond the scope of the present work (see e.g. \citet{Edw,Eve}),
to suppose first $h\ne0$, to construct the incremental ratio\[
\frac{f(x+h)-f(x)}{h}\]
 and, after suitable simplifications (sometimes using infinitesimal
properties), to take in the final result $h=0$. 
\item Fermat's method to find the maximum or minimum of a given function
$f(x)$ at $x=a$ was to take $e$ to be extremely small so that the
value of $f(x+h)$ was approximately equal to that of $f(x)$. In
modern, algebraic language, it can be said that $f(x+h)=f(x)$ only
if $h^{2}=0$, that is if $e$ is a first order infinitesimal. Fermat
was aware that this is not a {}``true'' equality but some kind of
approximation (ibidem). We will follow a similar idea to define $\ER$
introducing a suitable equivalence relation to represent this equality. 
\item Fermat has been described by \citet{EBell} as {}``the king of amateurs''
of mathematics, and hence we can suppose that in its mathematical
work the informal/intuitive part was stronger with respect to the
formal one. For this reason we can think that he would have liked
our idea to obtain a theory of infinitesimals preserving always the
intuitive meaning and without forcing the working mathematician to
be too much formal. 
\end{enumerate}
For these reason we chose the name {}``Fermat reals'' for our ring
$\ER$ (note: without the possessive case, to underline that we are
not attributing our construction of $\ER$ to Fermat).

\section{\label{sec:Definition-and-algebraic-prop-of-Fermat-reals}Definition
and algebraic properties of Fermat reals: The basic idea}

We start from the idea that a smooth ($\Cc^{\infty}$) function $f:\ER\freccia\ER$
is actually \emph{equal} to its tangent straight line in the first
order neighborhood e.g. of the point $x=0$, that is

\begin{equation}
\forall h\in D:\ f(h)=f(0)+h\cdot f'(0)\label{eq:FunctionEqualTangent}\end{equation}
 where $D$ is the subset of $\ER$ which defines the above-mentioned
neighborhood of $x=0$. The equality \eqref{eq:FunctionEqualTangent}
can be seen as a first-order Taylor's formula without remainder because
intuitively we think that $h^{2}=0$ for any $h\in D$ (indeed the
property $h^{2}=0$ defines the first order neighborhood of $x=0$
in $\ER$). These almost trivial considerations lead us to understand
many things: $\ER$ must necessarily be a ring and not a field because
in a field the equation $h^{2}=0$ implies $h=0$; moreover we will
surely have some limitation in the extension of some function from
$\R$ to $\ER$, e.g. the square root, because using this function
with the usual properties, once again the equation $h^{2}=0$ implies
$|h|=0$. On the other hand, we are also led to ask whether \eqref{eq:FunctionEqualTangent}
uniquely determines the derivative $f^{\prime}(0)$: because, even
if it is true that we cannot simplify by $h$, we know that the polynomial
coefficients of a Taylor's formula are unique in classical analysis.
In fact we will prove that

\begin{equation}
\exists!\, m\in\R\,\,\forall h\in D:\ f(h)=f(0)+h\cdot m\label{eq:IdeaDF}\end{equation}
 that is the slope of the tangent is uniquely determined in case it
is an ordinary real number. We will call formulas like \eqref{eq:IdeaDF}
\emph{derivation formula}s.

If we try to construct a model for \eqref{eq:IdeaDF} a natural idea
is to think our new numbers in $\ER$ as equivalence classes $[h]$
of usual functions $h:\R\freccia\R$. In this way we may hope both
to include the real field using classes generated by constant functions,
and that the class generated by $h(t)=t$ could be a first order infinitesimal
number. To understand how to define this equivalence relation we have
to think at \eqref{eq:FunctionEqualTangent} in the following sense:

\begin{equation}
f(h(t))\sim f(0)+h(t)\cdot f^{\prime}(0),\label{eq:IdeaFunction}\end{equation}
where the idea is that we are going to define $\sim$. If we think
$h(t)$ {}``sufficiently similar to $t$'', we can define $\sim$
so that (\ref{eq:IdeaFunction}) is equivalent to \[
\lim_{t\to0^{+}}\frac{f(h(t))-f(0)-h(t)\cdot f^{\prime}(0)}{t}=0,\]
 that is

\begin{equation}
x\sim y\DIff\lim_{t\to0^{+}}\frac{x_{t}-y_{t}}{t}=0.\label{eq:IdeaRelEq}\end{equation}
 In this way \eqref{eq:IdeaFunction} is very near to the definition
of differentiability for $f$ at 0. 

\noindent It is important to note that, because of de L'H$\hat{\text{o}}$pital's
theorem we have the isomorphism \[
\Cc^{1}(\R,\R)/\!\sim\,\,\,\simeq\,\,\R[x]/(x),\]
 the left hand side is (isomorphic to) the usual tangent bundle of
$\R$ and thus we obtain nothing new. It is not easy to understand
what set of functions we have to choose for $x$, $y$ in (\ref{eq:IdeaRelEq})
so as to obtain a non trivial structure. The first idea is to take
continuous functions at $t=0$, instead of more regular ones like
$\mathcal{C}^{1}$-functions, so that e.g. $h_{k}(t)=|t|^{1/k}$ becomes
a $k$-th order nilpotent infinitesimal ($h^{k+1}\sim0$); indeed
for almost all the results presented in this article, continuous functions
at $t=0$ work well. However, only in proving the non-trivial property 

\noindent \begin{equation}
\left(\forall x\in\ER:\ x\cdot f(x)=0\right)\then\forall x\in\ER:\ f(x)=0\label{eq:UniquenessIncrementalRatio}\end{equation}
we can see that it does not suffice to take continuous functions at
$t=0$. To prove \eqref{eq:UniquenessIncrementalRatio} the following
functions turned out to be very useful: 
\begin{defn}
\noindent \label{def:NilpotentFunctions}If $x:\R_{\ge0}\freccia\R$,
then we say that $x$ \emph{is nilpotent} iff $|x(t)-x(0)|^{k}=o(t)$
as $t\to0^{+}$, for some $k\in\N$. $\Nil$ will denote the set of
all the nilpotent functions.
\end{defn}
\noindent E.g. any Hoelder function $|x(t)-x(s)|\le c\cdot|t-s|^{\alpha}$
(for some constant $\alpha>0$) is nilpotent. The choice of nilpotent
functions instead of more regular ones establish a great difference
of our approach with respect to the classical definition of jets (see
e.g. \citet{Bro,Go-Gu}), that \eqref{eq:IdeaRelEq} may recall.

Another problem necessarily connected with the basic idea \eqref{eq:FunctionEqualTangent}
is that the use of nilpotent infinitesimals very frequently leads
to consider terms like $h_{1}^{i_{1}}\cdot\ldots\cdot h_{n}^{i_{n}}$.
For this type of products the first problem is to know whether $h_{1}^{i_{1}}\cdot\ldots\cdot h_{n}^{i_{n}}\ne0$
and what is the order $k$ of this new infinitesimal, that is for
what $k$ we have $(h_{1}^{i_{1}}\cdot\ldots\cdot h_{n}^{i_{n}})^{k}\ne0$
but $(h_{1}^{i_{1}}\cdot\ldots\cdot h_{n}^{i_{n}})^{k+1}=0$. We will
have a good frame if we will be able to solve these problems starting
from the order of each infinitesimal $h_{j}$ and from the values
of the powers $i_{j}\in\N$. On the other hand almost all the examples
of nilpotent infinitesimals are of the form $h(t)=t^{\alpha}$, with
$0<\alpha<1$, and their sums; these functions have also great properties
in the treatment of products of powers. It is for these reasons that
we shall focus our attention on the following family of functions
$x:\R_{\ge0}\freccia\R$ in the definition \eqref{eq:IdeaRelEq} of
$\sim$:
\begin{defn}
\label{def:LittleOhPolynomials} We say that $x$ \emph{is a little-oh
polynomial}, and we write $x\in\R_{o}[t]$ iff 
\begin{enumerate}
\item $x:\R_{\ge0}\freccia\R$ 
\item We can write\[
x_{t}=r+\sum\limits _{i=1}^{k}\alpha_{i}\cdot t^{a_{i}}+o(t)\quad\text{ as }\quad t\to0^{+}\]
 for suitable\[
k\in\N\]
\[
r,\alpha_{1},\dots,\alpha_{k}\in\R\]
\[
a_{1},\dots,a_{k}\in\R_{\ge0}\]

\end{enumerate}
\end{defn}
\noindent Hence a little-oh polynomial $x\in\R_{o}[t]$ is a polynomial
function with real coefficients, in the real variable $t\ge0$, with
generic positive powers of $t$, and up to a little-oh function as
$t\to0^{+}$. 
\begin{rem}
\label{rem:WithLittleOhWeMean}In the following, writing $x_{t}=y_{t}+o(t)$
as $t\to0+$ we will always mean\[
\lim_{t\to0^{+}}\frac{x_{t}-y_{t}}{t}=0\quad\text{and\ensuremath{\quad}}x_{0}=y_{0}.\]
 In other words, every little-oh function we will consider is continuous
as $t\to0^{+}$.\end{rem}
\begin{example*}
Simple examples of little-oh polynomials are the following:\end{example*}
\begin{enumerate}
\item $x_{t}=1+t+t^{1/2}+t^{1/3}+o(t)$ 
\item $x_{t}=r\quad\forall t$. Note that in this example we can take $k=0$,
and hence $\alpha$ and $a$ are the void sequence of reals, that
is the function $\alpha=a:\emptyset\freccia\R$, if we think of an
$n$-tuple $x$ of reals as a function $x:\left\{ 1,\dots,n\right\} \freccia\R$. 
\item $x_{t}=r+o(t)$ 
\end{enumerate}

\section{First properties of little-oh polynomials\label{sec:firstPropertiesOfLittle-ohPolynomials}}

\subsubsection*{Little-oh polynomials are nilpotent:}

First properties of little-oh polynomials are the following: if $x_{t}=r+\sum_{i=1}^{k}\alpha_{i}\cdot t^{a_{i}}+o_{1}(t)$
as $t\to0^{+}$ and $y_{t}=s+\sum_{j=1}^{N}\beta_{j}\cdot t^{b_{j}}+o_{2}(t)$,
then $(x+y)=r+s+\sum_{i=1}^{k}\alpha_{i}\cdot t^{a_{i}}+\sum_{j=1}^{N}\beta_{j}\cdot t^{b_{j}}+o_{3}(t)$
and $(x\cdot y)_{t}=rs+\sum_{i=1}^{k}s\alpha_{i}\cdot t^{a_{i}}+\sum_{j=1}^{N}r\beta_{j}\cdot t^{b_{i}}+\sum_{i=1}^{k}\sum_{j=1}^{N}\alpha_{i}\beta_{j}\cdot t^{a_{i}}t^{b_{j}}+o_{4}(t)$,
hence the set of little-oh polynomials is closed with respect to pointwise
sum and product. Moreover little-oh polynomials are nilpotent (see
Definition \ref{def:NilpotentFunctions}) functions; to prove this
we firstly prove that the set of nilpotent functions $\mathcal{N}$
is a subalgebra of the algebra $\R^{\R}$ of real valued functions.
Indeed, let $x$ and $y$ be two nilpotent functions such that $|x-x(0)|^{k}=o_{1}(t)$
and $|y-y(0)|^{N}=o_{2}(t)$, then we can write $x\cdot y-x(0)\cdot y(0)=x\cdot[y-y(0)]+y(0)\cdot[x-x(0)]$,
so that we can consider $|x\cdot[y-y(0)]|^{k}=|x|^{k}\cdot|y-y(0)|^{k}=|x|^{k}\cdot o_{1}(t)$
and $\frac{|x|^{k}\cdot o_{1}(t)}{t}\to0$ as $t\to0^{+}$ because
$|x|^{k}\to|x(0)|^{k}$, hence $x\cdot[y-y(0)]\in\mathcal{N}$. Analogously
$y(0)\cdot[x-x(0)]\in\mathcal{N}$ and hence the closure of $\mathcal{N}$
with respect to the product follows from the closure with respect
to the sum. The case of the sum follows from the following equalities
(where we use $x_{t}:=x(t)$, $u:=x-x_{0}$, $v:=y-y_{0}$, $|u_{t}|^{k}=o_{1}(t)$
and $|v_{t}|^{N}=o_{2}(t)$ and we have supposed $k\ge N$):\[
u^{k}=o_{1}(t),\,\,\, v^{k}=o_{2}(t)\]
 \[
(u+v)^{k}=\sum_{i=0}^{k}\binom{k}{i}u^{i}\cdot v^{k-i}\]
 \[
\forall i=0,\dots,k\pti\frac{u_{t}^{i}\cdot v_{t}^{k-i}}{t}=\frac{\left(u_{t}^{k}\right)^{\frac{i}{k}}\cdot\left(v_{t}^{k}\right)^{\frac{k-i}{k}}}{t^{\frac{i}{k}}\cdot t^{\frac{k-i}{k}}}=\left(\frac{u_{t}^{k}}{t}\right)^{\frac{i}{k}}\cdot\left(\frac{v_{t}^{k}}{t}\right)^{\frac{k-i}{k}}.\]
 Now we can prove that $\R_{o}[t]$ is a subalgebra of $\mathcal{N}$.
Indeed every constant $r\in\R$ and every power $t^{a_{i}}$ are elements
of $\mathcal{N}$ and hence $r+\sum_{i=1}^{k}\alpha_{i}\cdot t^{a_{i}}\in\mathcal{N}$,
so it remains to prove that if $y\in\mathcal{N}$ and $w=o(t)$, then
$y+w\in\mathcal{N}$, but this is a consequence of the fact that every
little-oh function is trivially nilpotent, and hence it follows from
the closure of $\mathcal{N}$ with respect to the sum.

\subsubsection*{Closure of little-oh polynomials with respect to smooth functions:\label{sub:ClosureOfLittle-ohPolyWRTSmoothFunctions}}

Now we want to prove that little-oh polynomials are preserved by smooth
functions, that is if $x\in\R_{o}[t]$ and $f:\R\freccia\R$ is smooth,
then $f\circ x\in\R_{o}[t]$. Let us fix some notations:\[
x_{t}=r+\sum_{i=1}^{k}\alpha_{i}\cdot t^{a_{i}}+w(t)\quad\text{with}\quad w(t)=o(t)\]
 \[
h(t):=x(t)-x(0)\quad\forall t\in\R_{\ge0}\]
 hence $x_{t}=x(0)+h_{t}=r+h_{t}$. The function $t\mapsto h(t)=\sum_{i=1}^{k}\alpha_{i}\cdot t^{a_{i}}+w(t)$
belongs to $\R_{o}[t]\subseteq\mathcal{N}$ so we can write $|h|^{N}=o(t)$
for some $N\in\N$ and as $t\to0^{+}$. From Taylor's formula we have\begin{align}
f(x_{t}) & =f(r+h_{t})=f(r)+\sum_{i=1}^{N}\frac{f^{(i)}(r)}{i!}\cdot h_{t}^{i}+f(x_{t})=f(r+h_{t})\label{eq:TaylorLittlePolyClosedSmooth}\\
 & =f(r)+\sum_{i=1}^{N}\frac{f^{(i)}(r)}{i!}\cdot h_{t}^{i}+o(h_{t}^{N})\end{align}
 But\[
\frac{|o(h_{t}^{N})|}{|t|}=\frac{|o(h_{t}^{N})|}{|h_{t}^{N}|}\cdot\frac{|h_{t}^{N}|}{|t|}\to0\]
 hence $o(h_{t}^{N})=o(t)\in\R_{o}[t]$. From this, the formula \eqref{eq:TaylorLittlePolyClosedSmooth},
the fact that $h\in\R_{o}[t]$ and using the closure of little-oh
polynomials with respect to ring operations, the conclusion $f\circ x\in\R_{o}[t]$
follows.

\section{Equality and decomposition of Fermat reals}
\begin{defn}
\label{def:equalityInFermatReals}Let $x$, $y\in\R_{o}[t]$, then
we say that $x\sim y$ or that $x=y$ in $\ER$ iff $x(t)=y(t)+o(t)$
as $t\to0^{+}$. Because it is easy to prove that $\sim$ is an equivalence
relation, we can define $\ER:=\R_{o}[t]/\sim$, i.e. $\ER$ is the
quotient set of $\R_{o}[t]$ with respect to the equivalence relation
$\sim$. 
\end{defn}
\noindent The equivalence relation $\sim$ is a congruence with respect
to pointwise operations, hence $\ER$ is a commutative ring. Where
it will be useful to simplify notations we will write {}``$x=y$
in $\ER$'' instead of $x\sim y$, and we will talk directly about
the elements of $\R_{o}[t]$ instead of their equivalence classes;
for example we can say that $x=y$ in $\ER$ and $z=w$ in $\ER$
imply $x+z=y+w$ in $\ER$.\\
 The immersion of $\R$ in $\ER$ is $r\longmapsto\hat{r}$ defined
by $\hat{r}(t):=r$, and in the sequel we will always identify $\hat{\R}$
with $\R$, which is hence a subring of $\ER$. Conversely if $x\in\ER$
then the map $\st(-):x\in\ER\mapsto\st x=x(0)\in\R$, which evaluates
each extended real in $0$, is well defined. We shall call $\st(-)$
the \emph{standard part map}. Let us also note that, as a vector space
over the field $\R$ we have $\dim_{\R}\ER=\infty$, and this underlines
even more the difference of our approach with respect to the classical
definition of jets. Our idea is instead more near to NSA, where standard
sets can be extended adding new infinitesimal points, and this is
not the point of view of jet theory.

With the following theorem we will introduce the decomposition of
a Fermat real $x\in\ER$, that is a unique notation for its standard
part and all its infinitesimal parts.
\begin{thm}
\label{thm:existenceUniquenessDecomposition}If $x\in\ER$, then there
exist one and only one sequence\[
(k,r,\alpha_{1},\ldots,\alpha_{k},a_{1},\ldots,a_{k})\]
 such that\[
k\in\N\]
\[
r,\alpha_{1},\dots,\alpha_{k},a_{1},\dots,a_{k}\in\R\]
and 
\begin{enumerate}
\item \label{enu:DecompositionSum}\emph{$x=r+\sum\limits _{i=1}^{k}\alpha_{i}\cdot t^{a_{i}}$}
in \emph{$\ER$} 
\item \emph{\label{enu:DecompositionIncreasingOfOrders}$0<a_{1}<a_{2}<\dots<a_{k}\le1$} 
\item \emph{\label{enu:DecompositionInfinitesimalPartsNotZero}$\alpha_{i}\ne0\quad\forall i=1,\dots,k$} 
\end{enumerate}
\end{thm}
\noindent In this statement we have also to include the void case
$k=0$ and $\alpha=a:\emptyset\freccia\R$. Obviously, as usual, we
use the definition $\sum_{i=1}^{0}b_{i}=0$ for the sum of an empty
set of numbers. As we shall see, this is the case where $x$ is a
standard real, i.e. $x\in\R$.\\
 In the following we will use the notations $t^{a}:=\diff{t_{1/a}}:=[t\in\R_{\ge0}\mapsto t^{a}\in\R]_{\sim}\in\ER$
so that e.g. $\diff{t_{2}}=t^{1/2}$ is a second order infinitesimal.
In general, as we will see from the definition of order of a generic
infinitesimal, $\diff{t_{a}}$ is an infinitesimal of order $a$.
In other words these two notations for the same object permit to emphasize
the difference between an actual infinitesimal $\diff{t_{a}}$ and
a potential infinitesimal $t^{1/a}$: an actual infinitesimal of order
$a\ge1$ corresponds to a potential infinitesimal of order $\frac{1}{a}\le1$
(with respect to the classical notion of order of an infinitesimal
function from calculus, see e.g. \citet{Pro,Sil}).
\begin{rem}
\label{rem:simplePropertiesOf_diff}Let us note that $\diff{t_{a}}\cdot\diff{t_{b}}=\diff{t_{\frac{ab}{a+b}}}$,
moreover $\diff{t}_{a}^{\alpha}:=(\diff{t_{a}})^{\alpha}=\diff{t_{\frac{a}{\alpha}}}$
for every $\alpha\ge1$ and finally $\diff{t_{a}}=0$ for every $a<1$.
E.g. $\diff{t}_{a}^{[a]+1}=0$ for every $a\in\R_{>0}$, where $[a]\in\N$
is the integer part of $a$, i.e. $[a]\le a<[a]+1$. 
\end{rem}
\noindent \textbf{Existence proof:}

Since $x\in\R_{o}[t]$, we can write $x_{t}=r+\sum_{i=1}^{k}\alpha_{i}\cdot t^{a_{i}}+o(t)$
as $t\to0^{+}$, where $r$, $\alpha_{i}\in\R$, $a_{i}\in\R_{\ge0}$
and $k\in\N$. Hence $x=r+\sum_{i=1}^{k}\alpha_{i}\cdot t^{a_{i}}$
in $\ER$ and our purpose is to pass from this representation of $x$
to another one that satisfies conditions \ref{enu:DecompositionSum},
\ref{enu:DecompositionIncreasingOfOrders} and \ref{enu:DecompositionInfinitesimalPartsNotZero}
of the statement. Since if $a_{i}>1$ then $\alpha_{i}\cdot t^{a_{i}}=0$
in $\ER$, we can suppose that $a_{i}\le1$ for every $i=1,\dots,k$.
Moreover we can also suppose $a_{i}>0$ for every $i$, because otherwise,
if $a_{i}=0$, we can replace $r\in\R$ by $r+\sum\{\alpha_{i}\,|\, a_{i}=0,\, i=1,\dots,k\}$.

\noindent Now we sum all the terms $t^{a_{i}}$ having the same $a_{i}$,
that is we can consider\[
\bar{\alpha_{i}}:=\sum\{\alpha_{j}\,|\, a_{j}=a_{i}\,,\, j=1,\dots,k\}\]
 so that in $\ER$ we have\[
x=r+\sum_{i\in I}\bar{\alpha_{i}}\cdot t^{a_{i}}\]
 where $I\subseteq\left\{ 1,\dots,k\right\} $, $\{a_{i}\,|\, i\in I\}=\{a,\dots,a_{k}\}$
and $a_{i}\ne a_{j}$ for any $i$, $j\in I$ with $i\ne j$. Neglecting
$\bar{\alpha_{i}}$ if $\bar{\alpha}_{i}=0$ and renaming $a_{i}$,
for $i\in I$, in such a way that $a_{i}<a_{j}$ if $i$, $j\in I$
with $i<j$, we obtain the existence result. Note that if $x=r\in\R$,
in the final step of this proof we have $I=\emptyset$.\medskip{}

\noindent \textbf{Uniqueness proof:}

Let us suppose that in $\ER$ we have\begin{equation}
x=r+\sum_{i=1}^{k}\alpha_{i}\cdot t^{a_{i}}=s+\sum_{j=1}^{N}\beta_{j}\cdot t^{b_{j}}\label{eq:uniquenessDecompositionHypothesis}\end{equation}
 where $\alpha_{i}$, $\beta_{j}$, $a_{i}$ and $b_{j}$ verify the
conditions of the statement. First of all $\st{x}=x(0)=r=s$ because
$a_{i}$, $b_{j}>0$. Hence $\alpha_{1}t^{a_{1}}-\beta_{1}t^{b_{1}}+\sum_{i}\alpha_{i}\cdot t^{a_{i}}-\sum_{j}\beta_{j}\cdot t^{b_{j}}=o(t)$.
By reduction to the absurd, if we had $a_{1}<b_{1}$, then collecting
the term $t^{a_{1}}$ we would have\begin{equation}
\alpha_{1}-\beta_{1}t^{b_{1}-a_{1}}+\sum_{i}\alpha_{i}\cdot t^{a_{i}-a_{1}}-\sum_{j}\beta_{j}\cdot t^{b_{j}-a_{1}}=\frac{o(t)}{t}\cdot t^{1-a_{1}}.\label{eq:firstTermUniquenessProofDecomposition}\end{equation}
 In \eqref{eq:firstTermUniquenessProofDecomposition} we have that
$\beta_{1}t^{b_{1}-a_{1}}\to0$ for $t\to0^{+}$ because $a_{1}<b_{1}$
by hypothesis; $\sum_{i}\alpha_{i}\cdot t^{a_{i}-a_{1}}\to0$ because
$a_{1}<a_{i}$ for $i=2,\dots,k$; $\sum_{j}\beta_{j}\cdot t^{b_{j}-a_{1}}\to0$
because $a_{1}<b_{1}<b_{j}$ for $j=2,\dots,N$, and finally $t^{1-a_{1}}$
is limited because $a_{1}\le1$. Hence for $t\to0^{+}$ we obtain
$\alpha_{1}=0$, which conflicts with condition \ref{enu:DecompositionInfinitesimalPartsNotZero}
of the statement. We can argue in a corresponding way if we had $b_{1}<a_{1}$.
In this way we see that we must have $a_{1}=b_{1}$. From this and
from equation \eqref{eq:firstTermUniquenessProofDecomposition} we
obtain\begin{equation}
\alpha_{1}-\beta_{1}+\sum_{i}\alpha_{i}\cdot t^{a_{i}-a_{1}}-\sum_{j}\beta_{j}\cdot t^{b_{j}-a_{1}}=\frac{o(t)}{t}\cdot t^{1-a_{1}}\label{eq:endFirstTerm}\end{equation}
 and hence for $t\to0^{+}$ we obtain $\alpha_{1}=\beta_{1}$. We
can now restart from \eqref{eq:endFirstTerm} to prove, in the same
way, that $a_{2}=b_{2}$, $\alpha_{2}=\beta_{2}$, etc. At the end
we must have $k=N$ because, otherwise, if we had e.g. $k<N$, at
the end of the previous recursive process, we would have\[
\sum_{j=k+1}^{N}\beta_{j}\cdot t^{b_{j}}=o(t).\]
 From this, collecting the terms containing $t^{b_{k+1}}$, we obtain\begin{equation}
t^{b_{k+1}-1}\cdot[\beta_{k+1}+\beta_{k+2}\cdot t^{b_{k+2}-b_{k+1}}+\dots+\beta_{N}\cdot t^{\beta_{N}-\beta_{k+1}}]\to0.\label{eq:uniquenessDecompositionN_equal_k}\end{equation}
 In this sum $\beta_{k+j}\cdot t^{b_{k+j}-b_{k+1}}\to0$ as $t\to0^{+}$,
because $b_{k+1}<b_{k+j}$ for $j>1$ and hence $\beta_{k+1}+\beta_{k+2}\cdot t^{b_{k+2}-b_{k+1}}+\dots+\beta_{N}\cdot t^{\beta_{N}-\beta_{k+1}}\to\beta_{k+1}\ne0$,
so from \eqref{eq:uniquenessDecompositionN_equal_k} we get $t^{b_{k+1}-1}\to0$,
that is $b_{k+1}>1$, in contradiction with the uniqueness hypothesis
$b_{k+1}\le1$.

Let us note explicitly that the uniqueness proof permits also to affirm
that the decomposition is well defined in $\ER$, i.e. that if $x=y$
in $\ER$, then the decomposition of $x$ and the decomposition of
$y$ are equal.$\qedNoNewLine$

On the basis of this theorem we introduce two notations: the first
one emphasizing the potential nature of an infinitesimal $x\in\ER$,
and the second one emphasizing its actual nature.
\begin{defn}
\label{def:potentialDecomposition}If $x\in\ER$, we say that\begin{equation}
x=r+\sum_{i=1}^{k}\alpha_{i}\cdot t^{a_{i}}\ \text{is the potential decomposition (of }x\text{)}\label{eq:potentialDecomposition}\end{equation}
 iff conditions \ref{enu:DecompositionSum}., \ref{enu:DecompositionIncreasingOfOrders}.,
and \ref{enu:DecompositionInfinitesimalPartsNotZero}. of Theorem
\ref{thm:existenceUniquenessDecomposition} are verified. Of course
it is implicit that the symbol of equality in \eqref{eq:potentialDecomposition}
has to be understood in $\ER$. 
\end{defn}
\noindent For example $x=1+t^{1/3}+t^{1/2}+t$ is a decomposition
because we have increasing powers of $t$. The only decomposition
of a standard real $r\in\R$ is the void one, i.e. that with $k=0$
and $\alpha=a:\emptyset\freccia\R$; indeed to see that this is the
case, it suffices to go along the existence proof again with this
case $x=r\in\R$ (or to prove it directly, e.g. by contradiction).
\begin{defn}
\label{def:actualDecomposition}Considering that $t^{a_{i}}=\diff{t_{1/a_{i}}}$
we can also use the following notation, emphasizing more the fact
that $x\in\ER$ is an actual infinitesimal:\begin{equation}
x=\st{x}+\sum_{i=1}^{k}\st{x_{i}}\cdot\diff{t_{b_{i}}}\label{eq:firstActualDecomposition}\end{equation}
 where we have used the notation $\st{x_{i}}:=\alpha_{i}$ and $b_{i}:=1/a_{i}$,
so that the condition that uniquely identifies all $b_{i}$ is $b_{1}>b_{2}>\dots>b_{k}\ge1$.
We call \eqref{eq:firstActualDecomposition} the \emph{actual decomposition}
of $x$ or simply the \emph{decomposition} of $x$. We will also use
the notation $\diff^{i}{x}:=\st{x_{i}}\cdot\diff{t_{b_{i}}}$ (and
simply $\diff{x}:=\diff{^{1}x}$) and we will call $\st{x_{i}}$ the
$i$\emph{-th standard part of} $x$ and $\diff{^{i}x}$ the $i$\emph{-th
infinitesimal part of} $x$ or the $i$-th \emph{differential} of
$x$. So let us note that we can also write\[
x=\st{x}+\sum_{i}\diff{^{i}x}\]
 and in this notation all the addenda are uniquely determined (the
number of them too). Finally, if $k\ge1$ that is if $x\in\ER\setminus\R$,
we set $\omega(x):=b_{1}$ and $\omega_{i}(x):=b_{i}$. The real number
$\omega(x)=b_{1}$ is the greatest order in the actual decomposition
\eqref{eq:firstActualDecomposition}, corresponding to the smallest
in the potential decomposition \eqref{eq:potentialDecomposition},
and is called the \emph{order} of the Fermat real $x\in\ER$. The
number $\omega_{i}(x)=b_{i}$ is called the $i$-th order of $x$.
If $x\in\R$ we set $\omega(x):=0$ and $\diff^{i}x:=0$. Observe
that in general $\omega(x)=\omega(\diff{x})$, $\diff{(\diff{x})}=\diff{x}$
and that, using the notations of the potential decomposition \eqref{def:potentialDecomposition},
we have $\omega(x)=1/a_{1}$. \end{defn}
\begin{example*}
If $x=1+t^{1/3}+t^{1/2}+t$, then $\st{x}=1$, $\diff{x}=\diff{t_{3}}$
and hence $x$ is a third order infinitesimal, i.e. $\omega(x)=3$,
$\diff{^{2}x}=\diff{t_{2}}$ and $\diff{^{3}x}=\diff{t}$; finally
all the standard parts are $\st{x_{i}}=1$. 
\end{example*}

\section{The ideals $D_{k}$}

In this section we will introduce the sets of nilpotent infinitesimals
corresponding to a $k$-th order neighborhood of 0. Every smooth function
restricted to this neighborhood becomes a polynomial of order $k$,
obviously given by its $k$-th order Taylor's formula (without remainder).
We start with a theorem characterizing infinitesimals of order less
than $k$.
\begin{thm}
\label{thm:characterizationOf_x_to_k_equal_0}If $x\in\ER$ and $k\in\N_{>1}$,
then $x^{k}=0$ in $\ER$ if and only if $\st{x}=0$ and $\omega(x)<k$. 
\end{thm}
\noindent \textbf{Proof:} If $x^{k}=0$, then taking the standard
part map of both sides, we have $\st{(x^{k})}=(\st{x})^{k}=0$ and
hence $\st{x}=0$. Moreover $x^{k}=0$ means $x_{t}^{k}=o(t)$ and
hence $\left(\frac{x_{t}}{t^{1/k}}\right)^{k}\to0$ and $\frac{x_{t}}{t^{1/k}}\to0$.
We rewrite this condition using the potential decomposition $x=\sum_{i=1}^{k}\alpha_{i}\cdot t^{a_{i}}$
of $x$ (note that in this way we have $\omega(x)=\frac{1}{a_{1}}$)
obtaining \[
\lim_{t\to0^{+}}\sum_{i}\alpha_{i}\cdot t^{a_{i}-\frac{1}{k}}=0=\lim_{t\to0^{+}}t^{a_{1}-\frac{1}{k}}\cdot\left[\alpha_{1}+\alpha_{2}\cdot t^{a_{2}-a_{1}}+\dots+\alpha_{k}\cdot t^{a_{k}-a_{1}}\right]\]
 But $\alpha_{1}+\alpha_{2}\cdot t^{a_{2}-a_{1}}+\dots+\alpha_{k}\cdot t^{a_{k}-a_{1}}\to\alpha_{1}\ne0$,
hence we must have that $t^{a_{1}-\frac{1}{k}}\to0$, and so $a_{1}>\frac{1}{k}$,
that is $\omega(x)<k$.\\
 Vice versa if $\st{x}=0$ and $\omega(x)<k$, then $x=\sum_{i=1}^{k}\alpha_{i}\cdot t^{a_{i}}+o(t)$,
and\[
\lim_{t\to0^{+}}\frac{x_{t}}{t^{1/k}}=\lim_{t\to0^{+}}\sum_{i}\alpha_{i}\cdot t^{a_{i}-\frac{1}{k}}+\lim_{t\to0^{+}}\frac{o(t)}{t}\cdot t^{1-\frac{1}{k}}\]
 But $t^{1-\frac{1}{k}}\to0$ because $k>1$ and $t^{a_{i}-\frac{1}{k}}\to0^{+}$
because $\frac{1}{a_{i}}\le\frac{1}{a_{1}}=\omega(x)<k$ and hence
$x^{k}=0$ in $\ER$.$\qedNoNewLine$

If we want that in a $k$-th order infinitesimal neighborhood a smooth
function is equal to its $k$-th Taylor's formula, we need to take
infinitesimals which are able to delete the remainder, that is, such
that $h^{k+1}=0$. The previous theorem permits to extend the definition
of the ideal $D_{k}$ to real number subscripts instead of natural
numbers $k$ only.
\begin{defn}
\label{def:idealD_a}If $a\in\R_{>0}\cup\left\{ \infty\right\} $,
then\[
D_{a}:=\left\{ x\in\ER\,|\,\st{x}=0,\ \omega(x)<a+1\right\} \]
Moreover, we will simply denote $D_{1}$ by $D$.\end{defn}
\begin{enumerate}
\item If $x=\diff{t_{3}}$, then $\omega(x)=3$ and $x\in D_{3}$. More
in general $\diff{t_{k}}\in D_{a}$ if and only if $\omega(\diff{t_{k}})=k<a+1$.
E.g. $\diff{t_{k}}\in D$ if and only if $1\le k<2$. 
\item $D_{\infty}=\bigcup_{a}D_{a}=\left\{ x\in\ER\,|\,\st{x}=0\right\} $
is the set of all the infinitesimals of $\ER$. 
\item $D_{0}=\left\{ 0\right\} $ because the only infinitesimal having
order strictly less than 1 is, by definition of order, $x=0$ (see
the Definition \ref{def:actualDecomposition}). 
\end{enumerate}
The following theorem gathers several expected properties of the sets
$D_{a}$ and of the order of an infinitesimal $\omega(x)$.
\begin{thm}
\label{thm:propertyOfIdeal_D_a}Let $a$, $b\in\R_{>0}$ and $x$,
$y\in D_{\infty}$, then 
\begin{enumerate}
\item \label{enu:IncreasingIdeals}$a\le b\then D_{a}\subseteq D_{b}$ 
\item \label{enu:xInItsOwnIdeal}$x\in D_{\omega(x)}$ 
\item \label{enu:consistencyWithIntegerDefinition}$a\in\N\then D_{a}=\{x\in\ER\,|\, x^{a+1}=0\}$ 
\item \label{enu:fromRealIdealToIntegerPower}$x\in D_{a}\then x^{\lceil a\rceil+1}=0$ 
\item \label{enu:integerPartOfOrder}$x\in D_{\infty}\setminus\{0\}$ and
$k=[\omega(x)]\then x\in D_{k}\setminus D_{k-1}$ 
\item \label{enu:diffOfProduct}$\diff{(x\cdot y)}=\diff{x}\cdot\diff{y}$ 
\item \label{enu:orderOfProduct}$x\cdot y\ne0\then{\displaystyle \frac{1}{\omega(x\cdot y)}=\frac{1}{\omega(x)}+\frac{1}{\omega(y)}}$ 
\item \label{enu:orderOfSum}$x+y\ne0\then\omega(x+y)=\omega(x)\vee\omega(y)$ 
\item \label{enu:D_aIsAnIdeal}$D_{a}$ is an ideal 
\end{enumerate}
\end{thm}
\noindent In this statement if $r\in\R$, then $\lceil r\rceil$ is
the \emph{ceiling} of the real $r$, i.e. the unique integer $\lceil r\rceil\in\mathbb{Z}$
such that $\lceil r\rceil-1<r\le\lceil r\rceil$. Moreover if $r$,
$s\in\R$, then $r\vee s:=\max(r,s)$.

Property \emph{\ref{enu:fromRealIdealToIntegerPower}}. of this theorem
cannot be proved substituting the ceiling $\lceil a\rceil$ with the
integer part $[a]$. In fact if $a=1.2$ and $x=\diff{t_{2.1}}$,
then $\omega(x)=2.1$ and $[a]+1=2$ so that $x^{[a]+1}=x^{2}=\diff{t_{\frac{2.1}{2}}}\ne0$
in $\ER$, whereas $\lceil a\rceil+1=3$ and $x^{3}=\diff{t_{\frac{2.1}{3}}}=0$.

Finally let us note the increasing sequence of ideals/neighborhoods
of zero:\begin{equation}
\{0\}=D_{0}\subset D=D_{1}\subset D_{2}\subset\dots\subset D_{k}\subset\dots\subset D_{\infty}.\label{eq:sequenceOfIdeals}\end{equation}
Because of \eqref{eq:sequenceOfIdeals} and of the property $\diff{t_{a}}=0$
if $a<1$, we can say that $\diff{t}$ is the smallest infinitesimals
and $\diff{t_{2}}$, $\diff{t_{3}}$, etc. are greater infinitesimals;
as we will see, this agree to corresponding order properties of these
infinitesimals.

\section{Products of powers of nilpotent infinitesimals}

In this section we will introduce some instruments that will be very
useful to decide whether a product of the form $h_{1}^{i_{1}}\cdot\ldots\cdot h_{n}^{i_{n}}$
, with $h_{k}\in D_{\infty}\setminus\{0\}$, is zero or whether it
belongs to some $D_{k}$. Generally speaking this problem is not trivial
in a ring (e.g. in SDG there is not an effective procedure to decide
this problem, see e.g. \citet{Lav}) and its solutions will be very
useful in the proofs of infinitesimal Taylor's formulas.
\begin{thm}
\label{thm:productOfPowers}Let $h_{1},\dots,h_{n}\in D_{\infty}\setminus\{0\}$
and $i_{1},\dots,i_{n}\in\N$, then 
\begin{enumerate}
\item \label{enu:iffForZeroProduct}${\displaystyle {\displaystyle h_{1}^{i_{1}}\cdot\ldots\cdot h_{n}^{i_{n}}=0}\quad\iff\quad\sum_{k=1}^{n}\frac{i_{k}}{\omega(h_{k})}>1}$ 
\item \label{enu:orderOfProductOfPowers}$h_{1}^{i_{1}}{\displaystyle \cdot\ldots\cdot h_{n}^{i_{n}}\ne0\then\frac{1}{\omega(h_{1}^{i_{1}}\cdot\ldots\cdot h_{n}^{i_{n}})}=\sum_{k=1}^{n}\frac{i_{k}}{\omega(h_{k})}}$ 
\end{enumerate}
\end{thm}
\noindent \textbf{Proof:} Let \begin{equation}
h_{k}=\sum_{r=1}^{N_{k}}\alpha_{kr}t^{a_{kr}}\label{eq:decompositionOf_h_k}\end{equation}
 be the potential decomposition of $h_{k}$ for $k=1,\dots,n$. Then
by Definition \ref{def:potentialDecomposition} of potential decomposition
and Definition \ref{def:actualDecomposition} of order, we have $0<a_{k1}<a_{k2}<\dots<a_{kN_{k}}\le1$
and $j_{k}:=\omega(h_{k})=\frac{1}{a_{k1}}$, hence $\frac{1}{j_{k}}\le a_{kr}$
for every $r=1,\dots,N_{k}$. Therefore from \eqref{eq:decompositionOf_h_k},
collecting the terms containing $t^{1/j_{k}}$ we have\[
h_{k}=t^{1/j_{k}}\cdot\left(\alpha_{k1}+\alpha_{k2}t^{a_{k2}-1/j_{k}}+\dots+\alpha_{kN_{k}}t^{a_{kN_{k}-1/j_{k}}}\right)\]
 and hence\begin{align}
h_{1}^{i_{1}}\cdot\ldots\cdot h_{n}^{i_{n}} & =t^{\frac{i_{1}}{j_{1}}+\dots+\frac{i_{n}}{j_{n}}}\cdot\left(\alpha_{11}+\alpha_{12}t^{a_{12}-\frac{1}{j_{1}}}+\dots+\alpha_{1N_{1}}t^{a_{1N_{1}}-\frac{1}{j_{1}}}\right)^{i_{1}}\cdot\ldots\nonumber \\
 & \ldots\cdot\left(\alpha_{n1}+\alpha_{n2}t^{a_{n2}-\frac{1}{j_{n}}}+\dots+\alpha_{nN_{n}}t^{a_{nN_{n}}-\frac{1}{j_{n}}}\right)^{i_{n}}\label{eq:productAfterGathering}\end{align}
 Hence if $\sum_{k}\frac{i_{k}}{j_{k}}>1$ we have that $t^{\frac{i_{1}}{j_{1}}+\dots+\frac{i_{n}}{j_{n}}}=0$
in $\ER$, so also $h_{1}^{i_{1}}\cdot\ldots\cdot h_{n}^{i_{n}}=0$.
Vice versa if $h_{1}^{i_{1}}\cdot\ldots\cdot h_{n}^{i_{n}}=0$, then
the right hand side of \eqref{eq:productAfterGathering} is a $o(t)$
as $t\to0^{+}$, that is\begin{align*}
t^{\frac{i_{1}}{j_{1}}+\dots+\frac{i_{n}}{j_{n}}-1}\cdot\left(\alpha_{11}+\alpha_{12}t^{a_{12}-\frac{1}{j_{1}}}+\dots+\alpha_{1N_{1}}t^{a_{1N_{1}}-\frac{1}{j_{1}}}\right)^{i_{1}}\cdot\ldots\\
\ldots\cdot\left(\alpha_{n1}+\alpha_{n2}t^{a_{n2}-\frac{1}{j_{n}}}+\dots+\alpha_{nN_{n}}t^{a_{nN_{n}}-\frac{1}{j_{n}}}\right)^{i_{n}} & \to0\end{align*}
 But each term $\left(\alpha_{k1}+\alpha_{k2}t^{a_{k2}-\frac{1}{j_{k}}}+\dots+\alpha_{kN_{k}}t^{a_{kN_{k}}-\frac{1}{j_{k}}}\right)^{i_{k}}\to\alpha_{k}^{i_{k}}\ne0$
so, necessarily, we must have $\frac{i_{1}}{j_{1}}+\dots+\frac{i_{n}}{j_{n}}-1>0$,
and this concludes the proof of \emph{\ref{enu:iffForZeroProduct}}.

\noindent To prove \emph{\ref{enu:orderOfProductOfPowers}}. it suffices
to apply recursively property \emph{\ref{enu:orderOfProduct}}. of
Theorem \ref{thm:propertyOfIdeal_D_a}.$\qedNoNewLine$
\begin{example}
\label{exa:productOfPowers}$\omega(\diff{t_{a_{1}}^{i_{1}}}\cdot\ldots\cdot\diff{t_{a_{n}}^{i_{n}}})^{-1}=\sum_{k}\frac{i_{k}}{\omega(\diff{t_{a_{k}}})}=\sum_{k}\frac{i_{k}}{a_{k}}$
and $\diff{t_{a_{1}}^{i_{1}}}\cdot\ldots\cdot\diff{t_{a_{n}}^{i_{n}}}=0$
if and only if $\sum_{k}\frac{i_{k}}{a_{k}}>1$, so e.g. $\diff{t}\cdot h=0$
for every $h\in D_{\infty}$. 
\end{example}
\noindent The following corollary gives a necessary and sufficient
condition to have $h_{1}^{i_{1}}\cdot\ldots\cdot h_{n}^{i_{n}}\in D_{p}\setminus\{0\}$.
\begin{cor}
\label{cor:idealForProductOfPowers}In the hypotheses of the previous
Theorem \ref{thm:productOfPowers} let $p\in\R_{>0}$, then we have\[
h_{1}^{i_{1}}\cdot\ldots\cdot h_{n}^{i_{n}}\in D_{p}\setminus\{0\}\quad\iff\quad\frac{1}{p+1}<\sum_{k=1}^{n}\frac{i_{k}}{\omega(h_{k})}\le1\]

\end{cor}
Let $h$, $k\in D$; because in this case $\sum_{k}\frac{i_{k}}{j_{k}+1}=\frac{1}{2}+\frac{1}{2}=1$
we always have\begin{equation}
h\cdot k=0.\label{eq:productOfFirstOrderInfinitesimalIsZero}\end{equation}
This is a great conceptual difference between Fermat reals and the
ring of SDG, where, not necessarily, the product of two first order
infinitesimal is zero. The consequences of this property of Fermat
reals arrive very deeply in the development of the theory of Fermat
reals, forcing us, e.g., to develop several new concepts if we want
to generalize the derivation formula \eqref{eq:IdeaDF} to functions
defined on infinitesimal domains, like $f:D\freccia\ER$ (see \citet{Gio4}).
We only mention here that looking at the simple Definition \ref{def:equalityInFermatReals},
the equality \eqref{eq:productOfFirstOrderInfinitesimalIsZero} has
an intuitively clear meaning, and it is to preserve this intuition
that we keep this equality instead of changing completely the theory
toward a less intuitive one.

\noindent Let us note explicitly that the possibility to prove these
results about products of powers of nilpotent infinitesimals is essentially
tied with the choice of little-oh polynomials in the definition of
the equivalence relation $\sim$ in Definition \ref{def:LittleOhPolynomials}.
Equally effective and useful results are not provable for the more
general family of nilpotent functions (see e.g. \citet{Gio3}).

\section{Identity principle for polynomials and invertible Fermat reals}

In this section we want to prove that if a polynomial $a_{0}+a_{1}x+a_{2}x^{2}+\dots+a_{n}x^{n}$
of $\ER$ is identically zero, then $a_{k}=0$ for all $k=0,\ldots,n$.
To prove this conclusion, it suffices to mean {}``identically zero''
as {}``equal to zero for every $x$ belonging to the extension of
an open subset of $\R$''. Therefore we firstly define what this
extension is.
\begin{defn}
\label{def:extensionOfSubsetsOfR}If $U$ is an open subset of $\R^{n}$,
then $\ext{U}:=\{x\in\ER^{n}\,|\,\st{x}\in U\}$. Here with the symbol
$\ext{\R^{n}}$ we mean $\ER^{n}:=\ER\times\ptind^{n}\times\ER$.
\end{defn}
The identity principle for polynomials can now be stated in the following
way and proved in standard manner using Vandermonde matrices.
\begin{thm}
\label{thm:PIP}Let $a_{0},\dots,a_{n}\in\ER$ and $U$ be an open
neighborhood of $0$ in $\R$ such that\begin{equation}
a_{0}+a_{1}x+a_{2}x^{2}+\dots+a_{n}x^{n}=0\quad\text{in }\ER\quad\forall x\in\ext{U}\label{eq:polyIdentity}\end{equation}

\noindent Then\[
a_{0}=a_{1}=\dots=a_{n}=0\quad\text{in }\ER\]

\end{thm}
Now, we want to see more formally that to prove \eqref{eq:FunctionEqualTangent}
we cannot embed the reals $\R$ into a field but only into a ring,
necessarily containing nilpotent element. In fact, applying \eqref{eq:FunctionEqualTangent}
to the function $f(h)=h^{2}$ for $h\in D$, where $D\subseteq\ER$
is a given subset of $\ER$, we have\[
f(h)=h^{2}=f(0)+h\cdot f'(0)=0\quad\forall h\in D.\]
Where we have supposed the preservation of the equality $f'(0)=0$
from $\R$ to $\ER$. In other words, if $D$ and $f(h)=h^{2}$ verify
\eqref{eq:FunctionEqualTangent}, then necessarily each element $h\in D$
must be a new type of number whose square is zero.

Because we cannot have property \eqref{eq:FunctionEqualTangent} and
a field at the same time, we need a sufficiently good family of cancellation
laws as substitutes. The simplest one of them is also useful to prove
the uniqueness of \eqref{eq:IdeaDF}:
\begin{thm}
\label{thm:firstCancellationLaw} If $x\in\ER$ is a Fermat real and
$r$, $s\in\R$ are standard real numbers, then\[
x\cdot r=x\cdot s\text{ in }\ER\quad\text{and}\quad x\ne0\then r=s\]

\end{thm}
\noindent \textbf{Proof:} From the Definition \ref{def:equalityInFermatReals}
of equality in $\ER$ and from $x\cdot r=x\cdot s$ we have\[
\lim_{t\to0^{+}}\frac{x_{t}\cdot(r-s)}{t}=0.\]
 But if we had $r\ne s$ this would implies $\lim_{t\to0^{+}}\frac{x_{t}}{t}=0$,
that is $x=0$ in $\ER$ and this contradicts the hypothesis $x\ne0$.$\qedNoNewLine$

The last result of this section takes its ideas from similar situations
of formal power series and gives also a formula to compute the inverse
of an invertible Fermat real.
\begin{thm}
\label{thm:formulaForInvertibleFermatdReals}Let $x=\st{x}+\sum_{i=1}^{n}\st{x_{i}}\cdot\diff{t_{a_{i}}}$
be the decomposition of a Fermat real $x\in\ER$. Then $x$ is invertible
if and only if $\st{x}\ne0$, and in this case\begin{equation}
\frac{1}{x}=\frac{1}{\st{x}}\cdot\sum_{j=0}^{+\infty}(-1)^{j}\cdot\left(\sum_{i=1}^{n}\frac{\st{x_{i}}}{\st{x}}\cdot\diff{t_{a_{i}}}\right)^{j}\label{eq:formulaforInvetibleFermatReal}\end{equation}

\end{thm}
\noindent In the formula \eqref{eq:formulaforInvetibleFermatReal}
we have to note that the series is actually a finite sum because any
$\diff{t_{a_{i}}}$ is nilpotent, e.g. $(1+\diff{t_{2}})^{-1}=1-\diff{t_{2}}+\diff{t_{2}^{2}}-\diff{t_{2}^{3}}+\dots=1-\diff{t_{2}}+\diff{t}$
because $\diff{t_{2}^{3}}=0$.

\textbf{Proof:} If $x\cdot y=1$ for some $y\in\ER$, then, taking
the standard parts of each side we have $\st{x}\cdot\st{y}=1$ and
hence $\st{x}\ne0$. Vice versa let $y:=\st{x}^{-1}\cdot\sum_{j=0}^{+\infty}(-1)^{j}\cdot\left(\sum_{i}\frac{\st{x_{i}}}{\st{x}}\diff{t_{a_{i}}}\right)^{j}$
and $h:=x-\st{x}=\sum_{i}\st{x_{i}}\diff{t_{a_{i}}}\in D_{\infty}$
so that we can also write\[
y=\st{x}^{-1}\cdot\sum_{j=0}^{+\infty}(-1)^{j}\cdot\frac{h^{j}}{\st{x}^{j}}\]
 But $h\in\ER$ is a little-oh polynomial with $h(0)=0$, so it is
also continuous, hence for a sufficiently small $\delta>0$ we have\[
\forall t\in(-\delta,\delta):\ \ \left|\frac{h_{t}}{\st{x}}\right|<1.\]
Therefore\[
\forall t\in(-\delta,\delta):\ \ y_{t}=\frac{1}{\st{x}}\cdot\left(1+\frac{h_{t}}{\st{x}}\right)^{-1}=\frac{1}{\st{x}+h_{t}}=\frac{1}{x_{t}}\]
 From this equality and from Definition \ref{def:equalityInFermatReals}
it follows $x\cdot y=1$ in $\ER$.$\qedNoNewLine$

\section{The derivation formula}

In this section we want to give a proof of \eqref{eq:IdeaDF} because
it has been the principal motivation for the construction of the ring
of Fermat reals $\ER$. Anyhow, before considering the proof of the
derivation formula, we have to extend a given smooth function $f:\R\freccia\R$
to a certain function $\ext f:\ER\freccia\ER$.
\begin{defn}
\label{def:extensionOfFunctions}Let $A$ be an open subset of $\R^{n}$,
$f:A\freccia\R$ a smooth function and $x\in\ext A$ then we define
\[
\ext f(x):=f\circ x.\]

\end{defn}
\noindent This definition is correct because we have seen that little-oh
polynomials are preserved by smooth functions, and because the function
$f$ is locally Lipschitz, so\[
\left|\frac{f(x_{t})-f(y_{t})}{t}\right|\le K\cdot\left|\frac{x_{t}-y_{t}}{t}\right|\quad\forall t\in(-\delta,\delta)\]
for a sufficiently small $\delta$ and some constant $K$, and hence
if $x=y$ in $\ER$, then also $\ext{f}(x)=\ext{f}(y)$ in $\ER$.

The function $\ext f$ is an extension of $f$, that is \[
\ext f(r)=f(r)\quad{\rm {in}\quad\ER\quad{\rm \forall r\in\R,}}\]
 as it follows directly from the definition of equality in $\ER$
(i.e. Definition \ref{def:equalityInFermatReals}), thus we can still
use the symbol $f(x)$ both for $x\in\ER$ and $x\in\R$ without confusion.
After the introduction of the extension of smooth functions, we can
also state the following useful \emph{elementary transfer theorem}
for equalities, whose proof follows directly from the previous definitions:
\begin{thm}
\label{thm:elementaryTransferTheorem}Let $A$ be an open subset of
$\R^{n}$, and $\tau$, $\sigma:A\freccia\R$ be smooth functions.
Then it results \[
\forall x\in{\ext A}\pti\ext\tau(x)=\ext\sigma(x)\]
 iff \[
\forall r\in A\pti\tau(r)=\sigma(r).\]

\end{thm}
\noindent Now we will prove the derivation formula \eqref{eq:IdeaDF}.
\begin{thm}
\label{thm:DerivationFormula} Let $A$ be an open set in $\R$, $x\in A$
and $f:A\freccia\R$ a smooth function\emph{,} then\begin{equation}
\exists!\, m\in\R\;\ \forall h\in D\pti f(x+h)=f(x)+h\cdot m.\label{eq:DerivationFormula}\end{equation}

\noindent In this case we have $m=f^{\prime}(x)$, where $f^{\prime}(x)$
is the usual derivative of $f$ at $x$. 
\end{thm}
\noindent \textbf{\textit{\emph{Proof:}}} \textit{\emph{Uniqueness
follows from the previous cancellation law Theorem \ref{thm:firstCancellationLaw},
indeed if $m_{1}\in\R$ and $m_{2}\in\R$ both verify \eqref{eq:DerivationFormula},
then $h\cdot m_{1}=h\cdot m_{2}$ for every $h\in D$. But there exists
a non zero first order infinitesimal, e.g. $\diff{t}\in D$, so from
Theorem \eqref{thm:firstCancellationLaw} it follows $m_{1}=m_{2}$.}}

To prove the existence part, take $h\in D$, so that $h^{2}=0$ in
$\ER$, i.e. $h_{t}^{2}=o(t)$ for $t\to0^{+}.$ But $f$ is smooth,
hence from its second order Taylor's formula we have\[
f(x+h_{t})=f(x)+h_{t}\cdot f'(x)+\frac{h_{t}^{2}}{2}\cdot f^{''}(x)+o(h_{t}^{2})\]
 But\[
\frac{o(h_{t}^{2})}{t}=\frac{o(h_{t}^{2})}{h_{t}^{2}}\cdot\frac{h_{t}^{2}}{t}\to0\quad\text{for }t\to0^{+}\]
 so\[
\frac{h_{t}^{2}}{2}\cdot f^{''}(x)+o(h_{t}^{2})=o_{1}(t)\quad\text{for }t\to0^{+}\]
 and we can write\[
f(x+h_{t})=f(x)+h_{t}\cdot f'(x)+o_{1}(t)\quad\text{for }t\to0^{+}\]
 that is \[
f(x+h)=f(x)+h\cdot f'(x)\quad\text{in }\ER\]
 and this proves the existence part because $f'(x)\in\R$.$\qedNoNewLine$

\noindent For example $e^{h}=1+h$, $\sin(h)=h$ and $\cos(h)=1$
for every $h\in D$.

Analogously we can prove the following infinitesimal Taylor's formula.
\begin{lem}
\label{thm:OrdinaryTaylorFor_nVariables}Let $A$ be an open set in
$\R^{d}$, $x\in A$, $n\in\N_{>0}$ and $f:A\freccia\R$ a smooth
function\emph{,} then\[
\forall h\in D_{n}^{d}\pti f(x+h)=\sum_{\substack{j\in\N^{d}\\
|j|\le n}
}\frac{h^{j}}{j!}\cdot\frac{\partial^{|j|}f}{\partial x^{j}}(x)\]

\end{lem}
\noindent For example $\sin(h)=h-\frac{h^{3}}{6}$ if $h\in D_{3}$
so that $h^{4}=0$.

It is possible to generalize several results of the present work to
functions of class $\mathcal{C}^{n}$ only, instead of smooth ones.
However it is an explicit purpose of this work to simplify statements
of results, definitions and notations, even if, as a result of this
searching for simplicity, its applicability will only hold for a more
restricted class of functions. Some more general results, stated for
$\mathcal{C}^{n}$ functions, but less simple can be found in \citet{Gio3}.

Note that $m=f^{\prime}(x)\in\R$, i.e. the slope is a standard real
number, and that we can use the previous formula with standard real
numbers $x$ only, and not with a generic $x\in\ER$, but we shall
remove this limitation in subsequent works (see also \citet{Gio4}).

If we apply this theorem to the smooth function $p(r):=\int_{x}^{x+r}f(t)\diff{t}$,
for $f$ smooth, then we immediately obtain the following result frequently
used in several informal calculations:
\begin{cor}
\label{cor:derivationFormulaForIntegrals_1D}Let $A$ be open in $\R$,
$x\in A$ and $f:A\freccia\R$ smooth. Then \[
\forall h\in D\pti\int_{x}^{x+h}f(t)\diff{t}=h\cdot f(x).\]
 Moreover $f(x)\in\R$ is uniquely determined by this equality. 
\end{cor}

\section{\label{SEC:INFINITESIMAL-AND-ORDER-PROPERTIES}Nilpotent infinitesimals
and order properties}

Like in other disciplines, also in mathematics the layout of a work
reflects the personal philosophical ideas of the authors. In particular
the present work is based on the idea that a good mathematical theory
is able to construct a good dialectic between formal properties, proved
in the theory, and their informal interpretations. The dialectic has
to be, as far as possible, in both directions: theorems proved in
the theory should have a clear and useful intuitive interpretation
and, on the other hand, the intuition corresponding to the theory
has to be able to suggest true sentences, i.e. conjectures or sketch
of proofs that can then be converted into rigorous proofs.

In a theory of new numbers, like the present one about Fermat reals,
the introduction of an order relation can be a hard test of the excellence
of this dialectic between formal properties and their informal interpretations.
Indeed if we introduce a new ring of numbers (like $\ER$) extending
the real field $\R$, we want that the new order relation, defined
on the new ring, will extend the standard one on $\R$. This extension
naturally leads to the wish of findings a geometrical representation
of the new numbers, according to the above principle of having a good
formal/informal dialectic.

We want to start this section showing that in our setting there is
a strong connection between some order properties and some algebraic
properties. In particular, we will show that it is not possible to
have good order properties and at the same time a uniqueness without
limitations in the derivation formula. In the following theorem we
can see that the property $h\cdot k=0$ is a general consequence if
we suppose to have a total order on $D$.
\begin{thm}
\noindent \label{thm:orderImpliesProductOfFirstOrderInfinitesimalsIsZero}Let
$(R,\le)$ be a generic ordered ring and $D\subseteq R$ a subset
of this ring, such that
\begin{enumerate}
\item $0\in D$ 
\item $\forall h\in D\pti h^{2}=0$ and $-h\in D$ 
\item $(D,\le)$ is a total order 
\end{enumerate}
\noindent then $h\cdot k=0$ for every $h$, $k\in D$.
\end{thm}
\noindent This theorem implies that if we want a total order in our
theory of infinitesimal numbers, and if in this theory we consider
$D=\{h\,|\, h^{2}=0\}$, then we must accept that the product of any
two elements of $D$ must be zero. For example, if we think that a
geometric representation of infinitesimals is not possible if we do
not have, at least, the trichotomy law, then in this theory we must
also have that the product of two first order infinitesimals is zero.

\noindent \textbf{Proof:} Let $h$, $k\in D$ be two elements of the
subset $D$. By hypotheses $0,$ $-h$, $-k\in D$, hence all these
elements are comparable with respect to the order relation $\le$,
because, by hypotheses this relation is total in $D$. E.g.\[
h\le k\quad\text{or}\quad k\le h\]
We will consider only the case $h\le k$, because analogously we can
deal with the case $k\le h$, simply exchanging everywhere $h$ with
$k$ and vice versa.

\noindent \emph{First sub-case}: \emph{$k\ge0$.\ \ \ }By multiplying
both sides of $h\le k$ by $k\ge0$ we obtain \begin{equation}
hk\le k^{2}\label{eq:hk_le_kSquare}\end{equation}
If $h\ge0$ then, multiplying by $k\ge0$ we have $0\le hk$, so from
\eqref{eq:hk_le_kSquare} we have $0\le hk\le k^{2}=0$, and hence
$hk=0$.

\noindent If $h\le0$ then, multiplying by $k\ge0$ we have\begin{equation}
hk\le0\label{eq:hk_le0}\end{equation}
If, furthermore, $h\ge-k$, then multiplying by $k\ge0$ we have $hk\ge-k^{2}$,
hence form \eqref{eq:hk_le0} $0\ge hk\ge-k^{2}=0$, hence $hk=0$.

\noindent If, otherwise, $h\le-k$, then multiplying by $-h\ge0$
we have $-h^{2}=0\le hk\le0$ from \eqref{eq:hk_le0}, hence $hk=0$.
This concludes the discussion of the case $k\ge0$.

\noindent \emph{Second sub-case}: $k\le0$.\ \ \ In this case we
have $h\le k\le0$. Multiplying both inequalities by $h\le0$ we obtain
$h^{2}=0\ge hk\ge0$ and hence $hk=0$.$\qedNoNewLine$

So, the trichotomy law is incompatible with the uniqueness in a possible
derivation formula like \begin{equation}
\exists!\, m\in R\pti\forall h\in D\pti f(h)=f(0)+h\cdot m\label{eq:derivationFormulaInAGenericRing}\end{equation}
framed in the ring $R$ of Theorem \ref{thm:orderImpliesProductOfFirstOrderInfinitesimalsIsZero}.
In fact, if $a$, $b\in D$ are two elements of the subset $D\subseteq R$,
then both $a$ and $b$ play the role of $m\in R$ in \eqref{eq:derivationFormulaInAGenericRing}
for the linear function\[
f:h\in D\mapsto h\cdot a=0\in R\]
So, if the derivation formula \eqref{eq:derivationFormulaInAGenericRing}
applies to linear functions (or less, to constant functions), the
uniqueness part of this formula cannot hold in the ring $R$.

In the next section we will introduce a natural and meaningful total
order relation on $\ER$. Therefore, the previous Theorem \ref{thm:orderImpliesProductOfFirstOrderInfinitesimalsIsZero}
strongly motivate that for the ring of Fermat reals $\ER$ we must
have that the product of two first order infinitesimals must be zero
and hence, that for the derivation formula in $\ER$ the uniqueness
cannot hold in its strongest form. Since we will also see that the
order relation permits to have a geometric representation of Fermat
reals, we can summarize the conclusions of this section saying that
the uniqueness in the derivation formula is incompatible with a natural
geometric interpretation of Fermat reals and hence with a good dialectic
between formal properties and informal interpretations in this theory.

\section{\label{sec:OrderRelation}Order relation}

From the previous sections one can draw the conclusion that the ring
of Fermat reals $\ER$ is essentially {}``the little-oh'' calculus.
But, on the other hand the Fermat reals give us more flexibility than
this calculus: working with $\ER$ we do not have to bother ourselves
with remainders made of {}``little-oh'', but we can neglect them
and use the powerful algebraic calculus with nilpotent infinitesimals.
But thinking the elements of $\ER$ as new numbers, and not simply
as {}``little-oh functions'', permits to treat them in a different
and new way, for example to define on them an order relation with
a clear geometrical interpretation.

First of all, let us introduce the useful notation\[
\forall^{0}t\ge0\pti\mathcal{P}(t)\]

\noindent and we will read the quantifier $\forall^{0}t\ge0$ saying
\emph{{}``for every $t\ge0$ (sufficiently) small}'', to indicate
that the property $\mathcal{P}(t)$ is true for all $t$ in some right
neighborhood of $t=0$ (recall that, by Definition \ref{def:LittleOhPolynomials},
our little-oh polynomials are always defined on $\R_{\ge0}$), i.e.\[
\exists\delta>0\pti\forall t\in[0,\delta)\pti\mathcal{P}(t).\]

\noindent The first heuristic idea to define an order relation is
the following\[
x\le y\iff x-y\le0\iff\exists z\pti z=0\text{ in }\ER\quad\text{and}\quad x-y\le z\]
More formally: 
\begin{defn}
\noindent \label{def:orderRelation}Let $x$, $y\in\ER$, then we
say\[
x\le y\]

\noindent iff we can find $z\in\ER$ such that $z=0$ in $\ER$ and\[
\forall^{0}t\ge0\pti x_{t}\le y_{t}+z_{t}\]

\end{defn}
\noindent Recall that $z=0$ in $\ER$ is equivalent to $z_{t}=o(t)$
for $t\to0^{+}$. It is immediate to see that we can equivalently
define $x\le y$ if and only if we can find $x'=x$ and $y'=y$ in
$\ER$ such that $x_{t}\le y_{t}$ for every $t$ sufficiently small.
From this it also follows that the relation $\le$ is well defined
on $\ER$, i.e. if $x'=x$ and $y'=y$ in $\ER$ and $x\le y$, then
$x'\le y'$ (recall that, to simplify the notations, we do not use
equivalence classes as elements of $\ER$ but directly little-oh polynomials).
As usual we will use the notation $x<y$ for $x\le y$ and $x\ne y$.
\begin{thm}
\noindent The relation $\le$ is an order, i.e. is reflexive, transitive
and anti-symmetric; it extends the order relation of $\R$ and with
it $(\ER,\le)$ is an ordered ring. Finally the following sentences
are equivalent:
\begin{enumerate}
\item $h\in D_{\infty}$, i.e. $h$ is an infinitesimal 
\item $\forall r\in\R_{>0}\pti-r<h<r$ 
\end{enumerate}
\end{thm}
\noindent Hence an infinitesimal can be thought of as a number with
standard part zero, or as a number smaller than every standard positive
real number and greater than every standard negative real number.

\noindent \textbf{Proof:} We only prove the prove the property\[
x\le y\quad\text{and}\quad w\ge0\then x\cdot w\le y\cdot w,\]
the others being a simple consequence of our Definition \ref{def:orderRelation}.
Let us suppose that\begin{align}
x_{t} & \le y_{t}+z_{t}\quad\forall^{0}t\ge0\label{eq:xLessThan_yPlus_z}\\
w_{t} & \ge z'_{t}\quad\forall^{0}t\ge0\nonumber \end{align}
 then $w_{t}-z'_{t}\ge0$ for every $t$ small and hence from \eqref{eq:xLessThan_yPlus_z}\[
x_{t}\cdot(w_{t}-z'_{t})\le y_{t}\cdot(w_{t}-z'_{t})+z_{t}\cdot(w_{t}-z'_{t})\quad\forall^{0}t\ge0\]
 from which it follows\[
x_{t}\cdot w_{t}\le y_{t}\cdot w_{t}+(-x_{t}z'_{t}-y_{t}z'_{t}+z_{t}w_{t}-z_{t}z'_{t})\quad\forall^{0}t\ge0\]
 But $-xz'-yz'+zw-zz'=0$ in $\ER$ because $z=0$ and $z'=0$ and
hence the conclusion follows.$\qedNoNewLine$ 
\begin{example*}
We have e.g. $\diff{t}>0$ and $\diff{t_{2}}-3\diff{t}>0$ because
for $t\ge0$ sufficiently small $t^{1/2}>3t$ and hence\[
t^{1/2}-3t>0\quad\forall^{0}t\ge0.\]
 From examples like these ones we can guess that our little-oh polynomials
are always locally comparable with respect to pointwise order relation,
and this is the first step to prove that for our order relation the
trichotomy law holds. In the following statement we will use the notation
$\forall^{0}t>0:\mathcal{P}(t)$, that naturally means\[
\forall^{0}t\ge0\pti t\ne0\then\mathcal{P}(t)\]

\noindent where $\mathcal{P}(t)$ is a generic property depending
on $t$.\end{example*}
\begin{lem}
\label{lem:pointwiseComparison}Let $x$, $y\in\ER$, then
\begin{enumerate}
\item \label{enu:standardPartAreDifferent}$\st{x}<\st{y}\then\forall^{0}t\ge0\pti x_{t}<y_{t}$ 
\item \label{enu:standardPartAreEqual}If $\st{x}=\st{y}$, then\[
\left(\forall^{0}t>0\pti x_{t}<y_{t}\right)\ \ \text{or}\ \ \left(\forall^{0}t>0\pti x_{t}>y_{t}\right)\ \ \text{or}\ \ \left(x=y\text{ in }\ER\right)\]

\end{enumerate}
\end{lem}
\noindent \textbf{Proof:}

\noindent \emph{\ref{enu:standardPartAreDifferent}}.)\quad{}Let
us suppose that $\st{x}<\st{y}$, then the continuous function $t\ge0\mapsto y_{t}-x_{t}\in\R$
assumes the value $y_{0}-x_{0}>0$ hence is locally positive, i.e.\[
\forall^{0}t\ge0\pti x_{t}<y_{t}\]
 \emph{\ref{enu:standardPartAreEqual}}.)\quad{}Now let us suppose
that $\st{x}=\st{y}$, and introduce a notation for the potential
decompositions of $x$ and $y$ (see Definition \ref{def:potentialDecomposition}).
From the definition of equality in $\ER$, we can always write \begin{align*}
x_{t} & =\st{x}+\sum_{i=1}^{N}\alpha_{i}\cdot t^{a_{i}}+z_{t}\quad\forall t\ge0\\
y_{t} & =\st{y}+\sum_{j=1}^{M}\beta_{j}\cdot t^{b_{j}}+w_{t}\quad\forall t\ge0\end{align*}
 where $x=\st{x}+\sum_{i=1}^{N}\alpha_{i}\cdot t^{a_{i}}$ and $y=\st{y}+\sum_{j=1}^{M}\beta_{j}\cdot t^{b_{j}}$
are the potential decompositions of $x$ and $y$ (hence $0<\alpha_{i}<\alpha_{i+1}\le1$
and $0<\beta_{j}<\beta_{j+1}\le1$), whereas $w$ and $z$ are little-oh
polynomials such that $z_{t}=o(t)$ and $w_{t}=o(t)$ for $t\to0^{+}$.

\noindent \emph{Case}: $a_{1}<b_{1}$\ \ \ In this case the least
power in the two decompositions is $\alpha_{1}\cdot t^{a_{1}}$, and
hence we expect that the second alternative of the conclusion is the
true one if $\alpha_{1}>0$, otherwise the first alternative will
be the true one if $\alpha_{1}<0$ (recall that always $\alpha_{i}\ne0$
in a decomposition). Indeed, let us analyze, for $t>0$, the condition
$x_{t}<y_{t}$: the following formulae are all equivalent to it\[
\ \sum_{i=1}^{N}\alpha_{i}\cdot t^{a_{i}}<\sum_{j=1}^{N}\beta_{j}\cdot t^{b_{j}}+w_{t}-z_{t}\]
 \[
t^{a_{1}}\cdot\left[\alpha_{1}+\sum_{i=2}^{N}\alpha_{i}\cdot t^{a_{i}-a_{1}}\right]<\ t^{a_{1}}\cdot\left[\sum_{j=1}^{N}\beta_{j}\cdot t^{b_{j}-a_{1}}+(w_{t}-z_{t})\cdot t^{-a_{1}}\right]\]
 \[
\alpha_{1}+\sum_{i=2}^{N}\alpha_{i}\cdot t^{a_{i}-a_{1}}<\sum_{j=1}^{N}\beta_{j}\cdot t^{b_{j}-a_{1}}+(w_{t}-z_{t})\cdot t^{-a_{1}}.\]
 Therefore, let us consider the function\[
f(t):=\sum_{j=1}^{N}\beta_{j}\cdot t^{b_{j}-a_{1}}+(w_{t}-z_{t})\cdot t^{-a_{1}}-\alpha_{1}-\sum_{i=2}^{N}\alpha_{i}\cdot t^{a_{i}-a_{1}}\quad\forall t\ge0\]
 We can write\[
(w_{t}-z_{t})\cdot t^{-a_{1}}=\frac{w_{t}-z_{t}}{t}\cdot t^{1-a_{1}}\]
 and $\frac{w_{t}-z_{t}}{t}\to0$ as $t\to0^{+}$ because $w_{t}=o(t)$
and $z_{t}=o(t)$. Furthermore, $a_{1}\le1$ hence $t^{1-a_{1}}$
is bounded in a right neighborhood of $t=0$. Therefore, $(w_{t}-z_{t})\cdot t^{-a_{1}}\to0$
and the function $f$ is continuous at $t=0$ too, because $a_{i}<a_{i}$
and $a_{1}<b_{1}<b_{j}$. By continuity, the function $f$ is locally
strictly positive if and only if $f(0)=-\alpha_{1}>0$, hence\begin{align*}
\left(\forall^{0}t>0\pti x_{t}<y_{t}\right) & \iff\alpha_{1}<0\\
\left(\forall^{0}t>0\pti x_{t}>y_{t}\right) & \iff\alpha_{1}>0\end{align*}
\emph{Case}: $a_{1}>b_{1}$\quad{}We can argue in an analogous way
with $b_{1}$ and $\beta_{1}$ instead of $a_{1}$ and $\alpha_{1}$.

\noindent \emph{Case}: $a_{1}=b_{1}$\quad{}We shall exploit the
same idea used above and analyze the condition $x_{t}<y_{t}$. The
following are equivalent ways to express this condition\[
t^{a_{1}}\cdot\left[\alpha_{1}+\sum_{i=2}^{N}\alpha_{i}\cdot t^{a_{i}-a_{1}}\right]<t^{a_{1}}\cdot\left[\beta_{1}+\sum_{j=2}^{N}\beta_{j}\cdot t^{b_{j}-a_{1}}+(w_{t}-z_{t})\cdot t^{-a_{1}}\right]\]
 \[
\alpha_{1}+\sum_{i=2}^{N}\alpha_{i}\cdot t^{a_{i}-a_{1}}<\beta_{1}+\sum_{j=2}^{N}\beta_{j}\cdot t^{b_{j}-a_{1}}+(w_{t}-z_{t})\cdot t^{-a_{1}}\]
 Hence, exactly as we have demonstrated above, we can state that\begin{align*}
\alpha_{1}<\beta_{1} & \then\forall^{0}t>0\pti x_{t}<y_{t}\\
\alpha_{1}>\beta_{1} & \then\forall^{0}t>0\pti x_{t}>y_{t}\end{align*}

\noindent Otherwise $\alpha_{1}=\beta_{1}$ and we can restart with
the same reasoning using $a_{2}$, $b_{2}$, $\alpha_{2}$, $\beta_{2}$,
etc. If $N=M$, the number of addends in the decompositions, using
this procedure we can prove that\[
\forall t\ge0\pti x_{t}=y_{t}+w_{t}-z_{t},\]
 that is $x=y$ in $\ER$.

It remains to consider the case, e.g., $N<M$. In this hypotheses,
using the previous procedure we would arrive at the following analysis
of the condition $x_{t}<y_{t}$: \[
0<\sum_{j>N}\beta_{j}\cdot t^{b_{j}}+w_{t}-z_{t}\]
 \[
0<t^{b_{N+1}}\cdot\Bigg[\beta_{N+1}+\sum_{j>N+1}\beta_{j}\cdot t^{b_{j}-b_{N+1}}+(w_{t}-z_{t})\cdot t^{-b_{N+1}}\Bigg]\]
 \[
0<\beta_{N+1}+\sum_{j>N+1}\beta_{j}\cdot t^{b_{j}-}{}^{b_{N+1}}+(w_{t}-z_{t})\cdot t^{-b_{N+1}}\]
 Hence\[
\beta_{N+1}>0\then\forall^{0}t>0\pti x_{t}<y_{t}\]

\[
\beta_{N+1}<0\then\forall^{0}t>0\pti x_{t}>y_{t}\]
 $\qedWithFinalEq$

\noindent This lemma can be used to find an equivalent formulation
of the order relation.
\begin{thm}
\noindent \label{thm:equivalentFormulationForOrderRelation}Let $x$,
$y\in\ER$, then
\begin{enumerate}
\item \label{enu:equivalentFormulationFor_x_le_y}$x\le y\iff\left(\forall^{0}t>0\pti x_{t}<y_{t}\right)$
~~or~~ $\ \ (x=y$ in $\ER)$ 
\item \label{enu:equivalentFormulationFor_x_less_y}$x<y\iff\left(\forall^{0}t>0\pti x_{t}<y_{t}\right)$
~~and~~ $(x\ne y$ in $\ER)$ 
\end{enumerate}
\end{thm}
\noindent \textbf{Proof:}

\noindent \emph{\ref{enu:equivalentFormulationFor_x_le_y}}.) $\Rightarrow$\ \ \ If
$\st{x}<\st{y}$ then, from the previous Lemma \ref{lem:pointwiseComparison}
we can derive that the first alternative is true. If $\st{x}=\st{y}$,
then from Lemma \ref{lem:pointwiseComparison} we have\begin{equation}
\left(\forall^{0}t>0\pti x_{t}<y_{t}\right)\quad\text{or}\quad\left(x=y\text{ in }\ER\right)\quad\text{or}\quad\left(\forall^{0}t>0\pti x_{t}>y_{t}\right)\label{eq:threeAlternatives}\end{equation}
 In the first two cases we have the conclusion. In the third case,
from $x\le y$ we obtain\begin{equation}
\forall^{0}t\ge0\pti x_{t}\le y_{t}+z_{t}\label{eq:x_le_yWith_z}\end{equation}
 with $z_{t}=o(t)$. Hence from the third alternative of \eqref{eq:threeAlternatives}
we have\[
0<x_{t}-y_{t}\le z_{t}\quad\forall^{0}t>0\]
 and hence $\lim_{t\to0^{+}}\frac{x_{t}-y_{t}}{t}=0$, i.e. $x=y$
in $\ER$.

\noindent \emph{\ref{enu:equivalentFormulationFor_x_le_y}}.) $\Leftarrow$\quad{}This
follows immediately from the reflexive property of $\le$ or from
the Definition \ref{def:orderRelation}.

\noindent \emph{\ref{enu:equivalentFormulationFor_x_less_y}}.) $\Rightarrow$\quad{}From
$x<y$ we have $x\le y$ and $x\ne y$, so the conclusion follows
from the previous \emph{\ref{enu:equivalentFormulationFor_x_le_y}}.

\noindent \emph{\ref{enu:equivalentFormulationFor_x_less_y}}.) \negthinspace{}$\Leftarrow$\quad{}From
$\forall^{0}t>0:x_{t}<y_{t}$ and from \emph{\ref{enu:equivalentFormulationFor_x_le_y}}.
it follows $x\le y$ and hence $x<y$ from the hypotheses $x\ne y$.$\qedNoNewLine$

\noindent Now we can prove that our order is total
\begin{cor}
\noindent Let $x$, $y\in\ER$, then in $\ER$ we have
\begin{enumerate}
\item \label{enu:trichotomyWithLessOrEqual}$x\le y\quad\text{or}\quad y\le x\quad\text{or}\quad x=y$ 
\item \label{enu:trichotomyWithLess}$x<y\quad\text{or}\quad y<x\quad\text{or}\quad x=y$ 
\end{enumerate}
\end{cor}
\noindent \textbf{Proof:}

\noindent \emph{\ref{enu:trichotomyWithLessOrEqual}}.)\quad{}If
$\st{x}<\st{y}$, then from Lemma \ref{lem:pointwiseComparison} we
have $x_{t}<y_{t}$ for $t\ge0$ sufficiently small. Hence from Theorem
\ref{thm:equivalentFormulationForOrderRelation} we have $x\le y$.
We can argue in the same way if $\st{x}>\st{y}$. Also the case $\st{x}=\st{y}$
can be handled in the same way using \emph{\ref{enu:standardPartAreEqual}}.
of Lemma \ref{lem:pointwiseComparison}.

\noindent \emph{\ref{enu:trichotomyWithLess}}.)\quad{}This part
is a general consequence of the previous one.$\qedNoNewLine$

From the proof of Lemma \ref{lem:pointwiseComparison} and from Theorem
\ref{thm:equivalentFormulationForOrderRelation} we can deduce the
following 
\begin{thm}
\label{thm:effectiveCriterionForTheOrder}Let $x$, $y\in\ER$. If
$\st{x}\ne\st{y}$, then\[
x<y\iff\st{x}<\st{y}\]

\noindent Otherwise, if $\st{x}=\st{y}$, then
\begin{enumerate}
\item If $\omega(x)>\omega(y)$, then $x>y$ iff $\st{x_{1}}>0$ 
\item If $\omega(x)=\omega(y)$, then\begin{align*}
\st{x_{1}}>\st{y_{1}} & \then x>y\\
\st{x_{1}}<\st{y_{1}} & \then x<y\end{align*}

\end{enumerate}
\end{thm}
\begin{example*}
The previous Theorem gives an effective criterion to decide whether
$x<y$ or not. Indeed, if the two standard parts are different, then
the order relation can be decided on the basis of these standard parts
only. E.g. $2+\diff{t_{2}}>3\diff{t}$ and $1+\diff{t_{2}}<3+\diff{t}$.
\end{example*}
\noindent Otherwise, if the standard parts are equal, we firstly have
to look at the order and at the first standard parts, i.e. $\st{x_{1}}$
and $\st{y_{1}}$, which are the coefficients of the biggest infinitesimals
in the decompositions of $x$ and $y$. E.g. $3\diff{t_{2}}>5\diff{t}$,
and $\diff{t_{2}}>a\diff{t}$ for every $a\in\R$, and $\diff{t}<\diff{t_{2}}<\diff{t_{3}}<\ldots<\diff{t_{k}}$
for every $k>3$, and $\diff{t_{k}}>0$.

\noindent If the orders are equal we have to compare the first standard
parts. E.g. $3\diff{t_{5}}>2\diff{t_{5}}$.

\noindent The other cases fall within the previous ones, because of
the properties of the ordered ring $\ER$. E.g. we have that $\diff{t_{5}}-2\diff{t_{3}}+3\diff{t}<\diff{t_{5}}-2\diff{t_{3}}+\diff{t_{3/2}}$
if and only if $3\diff{t}<\diff{t_{3/2}}$, which is true because
$\omega(\diff{t})=1<\omega(\diff{t_{3/2}})=\frac{3}{2}$. Finally
$\diff{t_{5}}-2\diff{t_{3}}+3\diff{t}>\diff{t_{5}}-2\diff{t_{3}}-\diff{t}$
because $3\diff{t}>-\diff{t}$.

\section{Absolute value, powers and logarithms}

Having a total order we can define the absolute value in the usual
way, and, exactly like for the real field $\R$, we can prove the
usual properties of the absolute value. Moreover, also the following
cancellation law is provable. 
\begin{thm}
\noindent Let $h\in\ER\setminus\{0\}$ and $r$, $s\in\R$, then\[
|h|\cdot r\le|h|\cdot s\then r\le s\]

\end{thm}
\noindent \textbf{Proof:} In fact if $|h|\cdot r\le|h|\cdot s$ then
from Theorem \ref{thm:equivalentFormulationForOrderRelation} we obtain
that either\begin{equation}
\forall^{0}t>0\pti|h_{t}|\cdot r\le|h_{t}|\cdot s\label{eq:absOf_h_tTimes_rLessOrEqualAbs_h_tTimes_s}\end{equation}
 or $|h|\cdot r=|h|\cdot s$. But $h\ne0$ so\[
\left(\forall^{0}t>0\pti h_{t}>0\right)\quad\text{or}\quad\left(\forall^{0}t>0\pti h_{t}<0\right)\]
 hence we can always find a $\bar{t}>0$ such that $|h_{\bar{t}}|\ne0$
and to which \eqref{eq:absOf_h_tTimes_rLessOrEqualAbs_h_tTimes_s}
is applicable. Therefore, in the first case we must have $r\le s$.
In the second one we have\[
|h|\cdot r=|h|\cdot s\]
 but $h\ne0$, hence $|h|\ne0$ and so the conclusion follows from
Theorem \ref{thm:firstCancellationLaw}.$\qedNoNewLine$

Due to the presence of nilpotent elements in $\ER$, we cannot define
powers $x^{y}$ and logarithms $\log_{x}y$ without any limitation.
E.g. we cannot define the square root having the usual properties,
like\begin{align}
x\in\ER & \then\sqrt{x}\in\ER\label{eq:SquareRootTakesLittle-ohPolyInLittle-ohPoly}\\
x=y\text{ in }\ER & \then\sqrt{x}=\sqrt{y}\text{ in }\ER\label{eq:SquareRootIsWellDefined}\\
 & \sqrt{x^{2}}=|x|\nonumber \end{align}
 because they are incompatible with the existence of $h\in D$ such
that $h^{2}=0$, but $h\ne0$. Indeed, the general property stated
in the Subsection \ref{sub:ClosureOfLittle-ohPolyWRTSmoothFunctions}
permits to obtain a property like \eqref{eq:SquareRootTakesLittle-ohPolyInLittle-ohPoly}
(i.e. the closure of $\ER$ with respect to a given operation) only
for smooth functions. Moreover, the Definition \ref{def:extensionOfFunctions}
states that to obtain a well defined operation we need a locally Lipschitz
function. For these reasons, we will limit $x^{y}$ to $x>0$ and
$x$ invertible only, and $\log_{x}y$ to $x$, $y>0$ and both $x$,
$y$ invertible.
\begin{defn}
Let $x$, $y\in\ER$, with $x$ strictly positive and invertible,
then
\begin{enumerate}
\item $x^{y}:=[t\ge0\mapsto x_{t}^{y_{t}}]_{=\text{ in }\ER}$ 
\item If $y>0$ and $y$ is invertible, then $\log_{x}y:=[t\ge0\mapsto log_{x_{t}}y_{t}]_{=\text{ in }\ER}$ 
\end{enumerate}
\end{defn}
\noindent Because of Theorem \ref{thm:equivalentFormulationForOrderRelation}
from $x>0$ we have \[
\forall^{0}t>0\pti x_{t}>0\]
so that, exactly as we proved in Subsection \ref{sub:ClosureOfLittle-ohPolyWRTSmoothFunctions}
and in Definition \ref{def:extensionOfFunctions}, the previous operations
are well defined in $\ER$ because $\st{x}\ne0\ne\st{y}$. From the
elementary transfer theorem \ref{thm:elementaryTransferTheorem} the
usual properties follow. To prove the usual monotonicity properties,
it suffices to use Theorem \ref{thm:equivalentFormulationForOrderRelation}.

Finally, it can be useful to state here the \emph{elementary transfer
theorem for inequalities}, whose proof follows immediately from the
definition of $\le$ and from Theorem \ref{thm:equivalentFormulationForOrderRelation}: 
\begin{thm}
\label{thm:elementaryTransferTheoremForInequalities}Let $A$ be an
open subset of $\R^{n}$, and $\tau$, $\sigma:A\freccia\R$ be smooth
functions. Then \[
\forall x\in{\ext A}\pti\ext\tau(x)\le\ext\sigma(x)\]
 iff \[
\forall r\in A\pti\tau(r)\le\sigma(r).\]

\end{thm}

\section{\label{sec:drawingOfFermatReals}Geometrical representation of Fermat
reals}

At the beginning of this article we argued that one of the conducting
idea in the construction of Fermat reals is to maintain always a clear
intuitive meaning. More precisely, we always tried, and we will always
try, to keep a good dialectic between provable formal properties and
their intuitive meaning. In this direction we can see the possibility
to find a geometrical representation of Fermat reals.

\begin{figure}[h]
\includegraphics[scale=0.15]{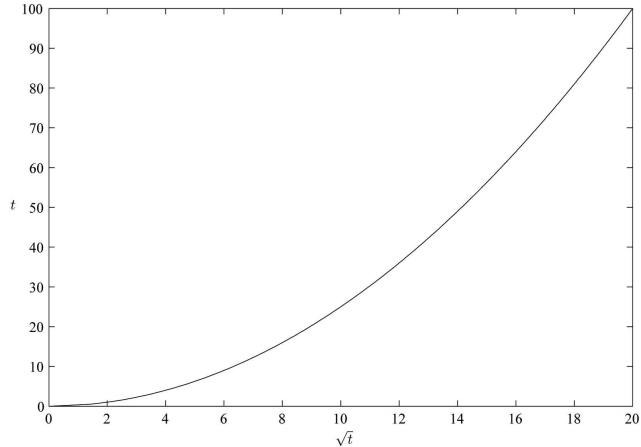} 

\caption{The function representing the Fermat real $\diff{t_{2}}\in D_{3}$}

\label{fig:representationOf_dt_2} 
\end{figure}

The idea is that to any Fermat real $x\in\ER$ we can associate the
function\begin{equation}
t\in\R_{\ge0}\mapsto\st{x}+\sum_{i=1}^{N}\st{x_{i}}\cdot t^{1/\omega_{i}(x)}\in\R\label{eq:functionsForGeometricalRepresentation}\end{equation}
 where $N$ is, of course, the number of addends in the decomposition
of $x$. Therefore, a geometric representation of this function is
also a geometric representation of the number $x$, because different
Fermat reals have different decompositions, see Theorem \ref{thm:existenceUniquenessDecomposition}.
Finally, we can guess that, because the notion of equality in $\ER$
depends only on the germ generated by each little-oh polynomial (see
Definition \ref{def:equalityInFermatReals}), we can represent each
$x\in\ER$ with only the first small part of the function \eqref{eq:functionsForGeometricalRepresentation}.
\begin{defn}
If $x\in\ER$ and $\delta\in\R_{>0}$, then\[
\text{\emph{graph}}_{\delta}(x):=\left\{ (\st{x}+\sum_{i=1}^{N}\st{x_{i}}\cdot t^{1/\omega_{i}(x)},t)\,|\,0\le t<\delta\right\} \]

\noindent where $N$ is the number of addends in the decomposition
of $x$. 
\end{defn}
\noindent Note that the value of the function are placed in the abscissa
position, so that the correct representation of $\text{graph}_{\delta}(x)$
is given by the figure \ref{fig:representationOf_dt_2}.

\noindent %

\noindent This inversion of abscissa and ordinate in the $\text{graph}_{\delta}(x)$
permits to represent this graph as a line tangent to the classical
straight line $\R$ and hence to have a better graphical picture.
Finally, note that if $x\in\R$ is a standard real, then $N=0$ and
the $\text{graph}_{\delta}(x)$ is a vertical line passing through
$\st{x}=x$.

\noindent The following theorem permits to represent geometrically
the Fermat reals
\begin{thm}
\noindent \label{thm:representationTheorem}If $\delta\in\R_{>0}$,
then the function \[
x\in\ER\mapsto\text{\emph{graph}}_{\delta}(x)\subset\R^{2}\]
 is injective. Moreover if $x$, $y\in\ER$, then we can find $\delta\in\R_{>0}$
(depending on $x$ and $y$) such that\[
x<y\]
 if and only if\begin{equation}
\forall p,q,t\pti(p,t)\in\text{\emph{graph}}_{\delta}(x)\ ,\ (q,t)\in\text{\emph{graph}}_{\delta}(y)\then p<q\label{eq:orderInTheGeometricalRepresentation}\end{equation}

\end{thm}
\noindent \textbf{Proof:} The application $\rho(x):=\text{graph}_{\delta}(x)$
for $x\in\ER$ is well defined because it depends on the terms $\st{x}$,
$\st{x_{i}}$ and $\omega_{i}(x)$ of the decomposition of $x$ (see
Theorem \ref{thm:existenceUniquenessDecomposition} and Definition
\ref{def:actualDecomposition}). Now, suppose that $\text{graph}_{\delta}(x)=\text{graph}_{\delta}(y)$,
then\begin{equation}
\forall t\in[0,\delta)\pti\st{x}+\sum_{i=1}^{N}\st{x_{i}}\cdot t^{1/\omega_{i}(x)}=\st{y}+\sum_{j=1}^{M}\st{y_{j}}\cdot t^{1/\omega_{j}(y)}.\label{eq:representingFunctionsAreLocallyEqual}\end{equation}
 Let us consider the Fermat reals generated by these functions, i.e.\begin{align*}
x': & =\left[t\ge0\mapsto\st{x}+\sum_{i=1}^{N}\st{x_{i}}\cdot t^{1/\omega_{i}(x)}\right]_{=\text{ in }\ER}\\
y': & =\left[t\ge0\mapsto\st{y}+\sum_{j=1}^{M}\st{y_{j}}\cdot t^{1/\omega_{j}(y)}\right]_{=\text{ in }\ER}\end{align*}
 then the decompositions of $x'$ and $y'$ are exactly the decompositions
of $x$ and~$y$\begin{align}
x' & =\st{x}+\sum_{i=1}^{N}\st{x_{i}}\diff{t_{\omega_{i}(x)}}=x\label{eq:xPrimeEqualx}\\
y' & =\st{y}+\sum_{j=1}^{M}\st{y_{j}}\diff{t_{\omega_{j}(y)}}=y.\label{eq:yPrimeEqualy}\end{align}
 But from \eqref{eq:representingFunctionsAreLocallyEqual} it follows
$x'=y'$ in $\ER$, and hence also $x=y$ from \eqref{eq:xPrimeEqualx}
and \eqref{eq:yPrimeEqualy}.

\begin{figure}[h]
\includegraphics{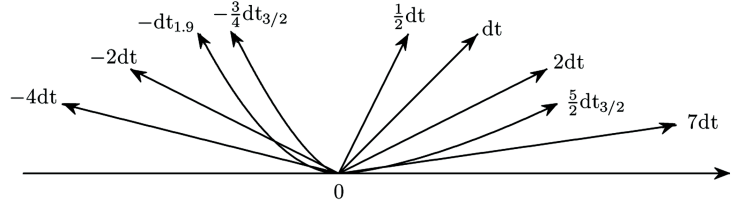} 

\caption{Some first order infinitesimals}

\label{fig:firstOrderInfinitesimals}
\end{figure}

Now suppose that $x<y$, then, using the same notations of the previous
part of this proof, we have also $x'=x$ and $y'=y$ and hence\[
x'=\st{x}+\sum_{i=1}^{N}\st{x_{i}}\cdot t^{1/\omega_{i}(x)}<\st{y}+\sum_{j=1}^{M}\st{y_{j}}\cdot t^{1/\omega_{j}(y)}=y'.\]
 We apply Theorem \ref{thm:equivalentFormulationForOrderRelation}
obtaining that locally $x'_{t}<y'_{t}$, i.e.\[
\exists\delta>0\pti\forall^{0}t\ge0\pti\st{x}+\sum_{i=1}^{N}\st{x_{i}}\cdot t^{1/\omega_{i}(x)}<\st{y}+\sum_{j=1}^{M}\st{y_{j}}\cdot t^{1/\omega_{j}(y)}.\]
 This is an equivalent formulation of \eqref{eq:orderInTheGeometricalRepresentation},
and, because of Theorem \ref{thm:equivalentFormulationForOrderRelation}
it is equivalent to $x'=x<y'=y$.$\qedNoNewLine$ 
\begin{example*}
In figure \ref{fig:firstOrderInfinitesimals} we have the representation
of some first order infinitesimals.
\end{example*}
\noindent %

\noindent The arrows are justified by the fact that the representing
function \eqref{eq:functionsForGeometricalRepresentation} is defined
on $\R_{\ge0}$ and hence has a clear first point and a direction.
The smaller is $\alpha\in(0,1)$ and the nearer is the representation
of the product $\alpha\diff{t}$, to the vertical line passing through
zero, which is the representation of the standard real $x=0$. Finally,
recall that $\diff{t_{k}}\in D$ if and only if $1\le k<2$.

\begin{figure}[h]
\includegraphics[scale=0.7]{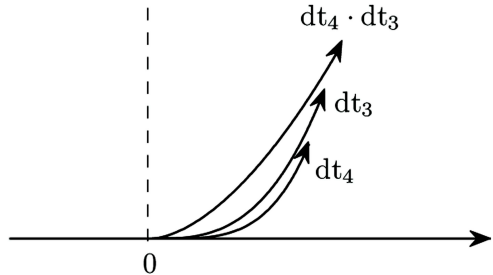} 

\caption{The product of two infinitesimals}

\label{fig:productOfTwoInfinitesimals}
\end{figure}

\noindent If we multiply two infinitesimals we obtain a smaller number,
hence one whose representation is nearer to the vertical line passing
through zero, as represented in figure \ref{fig:productOfTwoInfinitesimals}

\noindent %

\begin{figure}[h]
\includegraphics{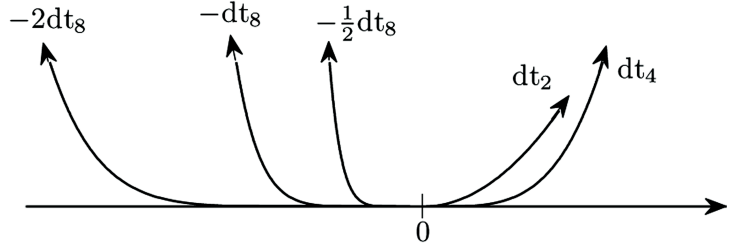} 

\caption{Some higher order infinitesimals}

\label{fig:higherOrderInfinitesimals}
\end{figure}

\noindent In figure \ref{fig:higherOrderInfinitesimals} we have a
representation of some infinitesimals of order greater than $1$.
We can see that the greater is the infinitesimal $h\in D_{a}$ (with
respect to the order relation $\le$ defined in $\ER$) and the higher
is the order of intersection of the corresponding line $\text{graph}_{\delta}(h)$.

\noindent %

\begin{figure}[h]
\includegraphics[scale=1.2]{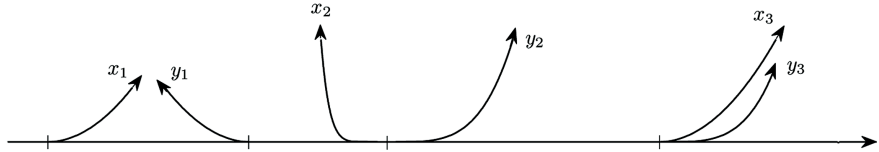} 

\caption{Different cases in which $x_{i}<y_{i}$}

\label{fig:representationOfOrderRelation}
\end{figure}

\noindent Finally, in figure \ref{fig:representationOfOrderRelation}
we represent the order relation on the basis of Theorem \ref{thm:representationTheorem}.

\noindent %

\noindent Intuitively, the method to see if $x<y$ is to look at a
suitably small neighborhood (i.e. at a suitably small $\delta>0$)
at $t=0$ of their representing lines $\text{graph}_{\delta}(x)$
and $\text{graph}_{\delta}(y)$: if, with respect to the horizontal
directed straight line, the curve $\text{graph}_{\delta}(x)$ comes
before the curve $\text{graph}_{\delta}(y)$, then $x$ is less than
$y$.

\section{\label{sec:someElementaryExample}Some elementary examples}

The elementary examples presented in this section want to show, in
a few rows, the simplicity of the algebraic calculus of nilpotent
infinitesimals. Here {}``simplicity'' means that the dialectic with
the corresponding informal calculations, used e.g. in engineering
or in physics, is really faithful. The importance of this dialectic
can be glimpsed both as a proof of the flexibility of the new language,
but also for researches in artificial intelligence like automatic
differentiation theories (see e.g. \citet{Gri} and references therein).
Last but not least, it may also be important for didactic or historical
researches. Several examples are directly taken from analogous of
\citet{Bel} and the reader is strongly invited to compare the two
theories in these cases. In particular, in our point of view, is not
positive, like in some parts of \citet{Bel}, to return back to a
non rigorous use of infinitesimals. Mathematical theories of infinitesimals,
like our ring of Fermat reals or NSA or SIA, are great opportunities
to avoid several fallacies of the informal approach (our discussion
in Section \ref{SEC:INFINITESIMAL-AND-ORDER-PROPERTIES} is a clear
example), and to advance further, with the new knowledge originating
from the rigorous theory, opening the possibility to use infinitesimal
methods in more general, and less intuitive, frameworks (like e.g.
infinite dimensional spaces of mappings, see \citet{Gio4}). Once
again, the key point is the dialectic between formal and informal
thoughts and not a single part only.

\subsection{\noindent The heat equation. }

\noindent In this and the following section we simply use the language
of $\ER$ to reformulate the corresponding deductions of \citet{Vla}.
Let us consider a body (identified with its localization) $B\subseteq\R^{3}$
and denote with $I_{B}:=\text{int}(B)$ its interior. On $I_{B}$
are given the smooth functions $\rho:I_{B}\freccia\R$, $c:I_{B}\freccia\R$
and $k:I_{B}\freccia\R$, interpreted respectively as the mass density,
the specific heat capacity and the coefficient of thermal conductivity.
Let us note that assuming these functions as defined on $I_{B}$ without
any favored direction corresponds physically to assume that $B$ is
an isotropous body. Moreover, let $u:I_{B}\times[0,+\infty)\freccia\R$
be the smooth function representing the temperature of the body $B$
at each point $x\in I_{B}$ and time $t\in[0,+\infty)$. To deduce
the heat diffusion equation we fix an internal point $x\in I_{B}$
and an infinitesimal volume $V$. More precisely, we say that a subset
of $\ER^{3}$ of the form \begin{equation}
V=V(x,\delta\underline{x})=\left\{ y\in\ER^{3}\,|\,-\delta x_{i}\le2(y-x)\cdot\vec{e}_{i}\le\delta x_{i}\quad\forall i=1,2,3\right\} \label{eq:infParallelepiped}\end{equation}
is an infinitesimal parallelepiped if $\delta v:=\delta x{}_{1}\cdot\delta x_{2}\cdot\delta x_{3}\in D_{\infty}$,
i.e. if the corresponding volume is an infinitesimal of some order.
Here $(\vec{e}_{1},\vec{e}_{2},\vec{e}_{3})$ is the natural base
of $\R^{3}$ and notations of the form $\delta y\in\ER$ are only
useful to underline that the infinitesimal increment is associated
to the variable $y$: here $\delta$ is not an operator and we use
it instead of the common $dy$ to avoid confusion with our $\diff{y}$
introduced in Definition \ref{def:actualDecomposition}. Because $x\in I_{B}$,
the inclusion $V\subseteq\ext{B}$ follows, so that $V$ can be thought
as the sub-body of $B$ corresponding to the infinitesimal parallelepiped
parallel to coordinate axis and centered at $x$. This sub-body $V$
interacts thermally with its complement $\mathcal{C}V:=\ext{B}\setminus V$
and with external sources of heat. In the infinitesimal time interval
$\delta t\in D_{\infty}$, the sub-body $V$ exchanges with its complement
$\mathcal{C}V$ the heat flowing perpendicularly to the surface of
$V$ (Fourier's law):\begin{equation}
Q_{\mathcal{C}V,V}=\delta t\cdot\sum_{i=1}^{3}\delta s_{i}\cdot\left[k(x+\delta\vec{h}_{i})\cdot\frac{\partial u}{\partial\vec{e}_{i}}(x+\delta\vec{h}_{i},t)-k(x-\delta\vec{h}_{i})\cdot\frac{\partial u}{\partial\vec{e}_{i}}(x-\delta\vec{h}_{i},t)\right],\label{eq:heatFromComplement}\end{equation}
where $\delta\vec{h}_{i}:=\frac{1}{2}\delta x_{i}\cdot\vec{e}_{i}\in\ER^{3}$
and $\delta s_{i}:=\prod_{j\ne i}\delta x_{j}\in\ER$. Choosing the
infinitesimals so that\[
\delta v\cdot\delta t\in D,\]
we have that $\delta t\cdot\delta s_{i}\cdot\left(\delta x_{i}\right)^{2}=\delta t\cdot\delta v\cdot\delta x_{i}=0$
from Theorem \ref{thm:productOfPowers} (e.g. we can choose $\delta x_{i}=\diff{t}_{6}$
and $\delta t=\diff{t}_{2}$). From this and the use of infinitesimal
Taylor's formula in \eqref{eq:heatFromComplement}, simple calculations
give\begin{equation}
Q_{\mathcal{C}V,V}=\text{div}\left[k\cdot\text{grad}(u)\right](x,t)\cdot\delta v\cdot\delta t.\label{eq:heatCV-afterGauss}\end{equation}
Of course, this calculations correspond to the infinitesimal version
of the Gauss-Ostrogradskij theorem. Interacting thermally with external
sources, the sub-body $V$ exchanges the heat\begin{equation}
Q_{\text{ext},V}=F(x,t)\cdot\delta v\cdot\delta t,\label{eq:heatFromExt}\end{equation}
where $F:I_{B}\times[0,+\infty)\freccia\R$ is a smooth function representing
the intensity of the thermal sources. The total heat $Q_{\mathcal{C}V,V}+Q_{\text{ext},V}$
corresponds to an increasing of temperature of $V$ equal to $u(x,t+\delta t)-u(x,t)$
and hence to an exchange of heat with the environment equal to\begin{equation}
Q_{\text{env},V}=\left[u(x,t+\delta t)-u(x,t)\right]\cdot c(x)\cdot\rho(x)\cdot\delta v=Q_{\mathcal{C}V,V}+Q_{\text{ext},V}.\label{eq:heatOf-V}\end{equation}
From this and \eqref{eq:heatCV-afterGauss}, \eqref{eq:heatFromExt},
the infinitesimal Taylor's formula and the cancellation law we obtain
the conclusion:\[
c(x)\cdot\rho(x)\cdot\frac{\partial u}{\partial t}(x,t)=\text{div}\left[k\cdot\text{grad}(u)\right](x,t)+F(x,t).\]
To stress that the previous deduction is now completely rigorous we
can now state the following theorem, without any mention to the physical
interpretation:
\begin{thm}
Let $B\subseteq\R^{d}$ and $I_{B}:=\text{\emph{int}}(B)$ its interior.
Let us consider the smooth functions $\rho:I_{B}\freccia\R$, $c:I_{B}\freccia\R$,
$k:I_{B}\freccia\R$, $u:I_{B}\times[0,+\infty)\freccia\R$ and $F:I_{B}\times[0,+\infty)\freccia\R$.
Finally let us consider a point $(x,t)\in I_{B}\times[0,+\infty)$
and define $V$, $Q_{\mathcal{C}V,V}$, $Q_{\text{\emph{ext}},V}$,
$Q_{\text{\emph{env}},V}$ as in \eqref{eq:infParallelepiped}, \eqref{eq:heatFromComplement},
\eqref{eq:heatFromExt} and \eqref{eq:heatOf-V}, where $\delta v\cdot\delta t\in D$.
Then it results\[
Q_{\text{\emph{env}},V}=Q_{\mathcal{C}V,V}+Q_{\text{\emph{ext}},V}\]
if and only if the following relation holds\[
c(x)\cdot\rho(x)\cdot\frac{\partial u}{\partial t}(x,t)=\text{\emph{div}}\left[k\cdot\text{\emph{grad}}(u)\right](x,t)+F(x,t).\]

\end{thm}
Unfortunately, this statement does not sufficiently underline the
great difference that takes place between the physical content in
the definition of $Q_{\mathcal{C}V,V}$, i.e. the Fourier's law, and
that in the definition of $Q_{\text{ext},V}$. In an axiomatic framework
for thermodynamics (see e.g. \citet{Tru}), the notion of heat flux
$Q_{AB}$ going from a body $A$ to a body $B$ can be taken as primitive;
in that case \eqref{eq:heatFromComplement} becomes an important assumption,
whereas \eqref{eq:heatFromExt} is simply the definition of the intensity
$F(x,t)=\frac{Q_{\text{ext},V}}{\delta v\cdot\delta t}$.

\subsection{Electric dipole.}

In elementary Physics, an electric dipole is usually defined as {}``\emph{a
pair of charges with opposite sign placed at a distance $d$ very
less than the distance $r$ from the observer''}. Conditions like
$r\gg d$ are frequently used in Physics and very often we obtain
a correct formalization if we ask $d\in\ER$ infinitesimal but $r\in\R\setminus\{0\}$,
i.e. $r$ finite. Thus we can define an electric dipole as a pair
$(p_{1},p_{2})$ of electric particles, with charges of equal intensity
but with opposite sign such that their mutual distance at every time
$t$ is a first order infinitesimal: \begin{equation}
\forall t\pti\vert p_{1}(t)-p_{2}(t)\vert=:\vert\vec{d}_{t}\vert=:d_{t}\in D.\label{eq:_DefinitionOfDipole}\end{equation}
 In this way we can calculate the potential at the point $x$ using
the properties of $D$ and using the hypothesis that $r$ is finite
and not zero. In fact we have \[
\varphi(x)=\frac{q}{4\pi\epsilon_{0}}\cdot\left(\frac{1}{r_{1}}-\frac{1}{r_{2}}\right)\qquad\qquad\vec{r_{i}}:=x-p_{i}\]
 and if $\vec{r}:=\vec{r}_{2}-\frac{\vec{d}}{2}$ then \[
\frac{1}{r_{2}}=\left(r^{2}+\frac{d^{2}}{4}+\vec{r}\boldsymbol{\cdot}\vec{d}\right)^{-1/2}=r^{-1}\cdot\left(1+\frac{\vec{r}\boldsymbol{\cdot}\vec{d}}{r^{2}}\right)^{-1/2}\]
 because for (\ref{eq:_DefinitionOfDipole}) $d^{2}=0$. For our hypotheses
on $d$ and $r$ we have that ${\displaystyle \frac{\vec{r}\boldsymbol{\cdot}\vec{d}}{r^{2}}\in D}$
hence from the derivation formula \[
\left(1+\frac{\vec{r}\boldsymbol{\cdot}\vec{d}}{r^{2}}\right)^{-1/2}=1-\frac{\vec{r}\boldsymbol{\cdot}\vec{d}}{2r^{2}}\]
 In the same way we can proceed for $1/r_{1}$, hence: \begin{align*}
\varphi(x) & =\frac{q}{4\pi\epsilon_{0}}\cdot\frac{1}{r}\cdot\left(1+\frac{\vec{r}\boldsymbol{\cdot}\vec{d}}{2r^{2}}-1+\frac{\vec{r}\boldsymbol{\cdot}\vec{d}}{2r^{2}}\right)=\\
 & =\frac{q}{4\pi\epsilon_{0}}\cdot\frac{\vec{r}\boldsymbol{\cdot}\vec{d}}{r^{3}}\end{align*}
 The property $d^{2}=0$ is also used in the calculus of the electric
field and for the moment of momentum.

\subsection{Newtonian limit in Relativity.}

Another example in which we can formalize a condition like $r\gg d$
using the previous ideas is the Newtonian limit in Relativity; in
it we can suppose to have 
\begin{itemize}
\item $\,\forall t\pti v_{t}\in D_{2}\quad\text{and}\quad c\in\R$ 
\item $\,\forall x\in M_{4}\pti g_{ij}(x)=\eta_{ij}+h_{ij}(x)\quad\text{with}\quad h_{ij}(x)\in D.$ 
\end{itemize}
where $\left(\eta_{ij}\right)_{ij}$ is the matrix of the Minkowski's
metric. This conditions can be interpreted as $v_{t}\ll c$ and $h_{ij}(x)\ll1$
(low speed with respect to the speed of light and weak gravitational
field). In this way we have, e.g. the equalities: \[
\frac{1}{\sqrt{{\displaystyle 1-\frac{v^{2}}{c^{2}}}}}=1+\frac{v^{2}}{2c^{2}}\qquad\text{and}\qquad\sqrt{1-h_{44}(x)}=1-\frac{1}{2}\, h_{44}(x).\]

\subsection{Linear differential equations.}

Let \begin{gather*}
L(y):=A_{\sss0}\frac{\diff{}^{\sss N}y}{\diff{}t^{\sss N}}+\ldots+A_{\sss N-1}\frac{\diff{}y}{\diff{}t}+A_{\sss N}\cdot y=0\end{gather*}
 be a linear differential equation with constant coefficients. Once
again we want \emph{to discover} independent solutions in case the
characteristic polynomial has multiple roots e.g. \[
(r-r_{\sss1})^{2}\cdot(r-r_{\sss3})\cdot\ldots\cdot(r-r_{\sss N})=0.\]
 The idea is that in $\ER$ we have $(r-r_{1})^{2}=0$ also if $r=r_{1}+h$
with $h\in D$. Thus $y(t)={\rm e}^{(r_{1}+h)t}$ is a solution too.
But ${\rm e}^{(r_{1}+h)t}={\rm e}^{r_{1}t}+ht\cdot{\rm e}^{r_{1}t}$,
hence \begin{align*}
L\left[{\rm e}^{(r_{1}+h)t}\right] & =0\\
{} & =L\left[{\rm e}^{r_{1}t}+ht\cdot{\rm e}^{r_{1}t}\right]\\
{} & =L\left[{\rm e}^{r_{1}t}\right]+h\cdot L\left[t\cdot{\rm e}^{r_{1}t}\right]\end{align*}
 We obtain $L\left[t\cdot{\rm e}^{r_{1}t}\right]=0$, that is $y_{1}(t)=t\cdot{\rm e}^{r_{1}t}$
must be a solution. Using $k$-th order infinitesimals we can deal
with other multiple roots in a similar way.

\subsection{Circle of curvature.}

A simple application of the infinitesimal Taylor's formula is the
parametric equation for the circle of curvature, that is the circle
with second order osculation with a curve $\gamma:[0,1]\freccia\R^{3}$.
In fact if $r\in(0,1)$ and $\dot{\gamma}_{r}$ is a unit vector,
from the second order infinitesimal Taylor's formula we have \begin{equation}
\forall h\in D_{2}\pti\gamma(r+h)=\gamma_{r}+h\,\dot{\gamma}_{r}+\frac{h^{2}}{2}\,\ddot{\gamma}_{r}=\gamma_{r}+h\,\vec{t}_{r}+\frac{h^{2}}{2}c_{r}\,\vec{n}_{r}\label{eq:_CircleOfCurvature}\end{equation}
 where $\vec{n}$ is the unit normal vector, $\vec{t}$ is the tangent
one and $c_{r}$ the curvature. But once again from Taylor's formula
we have $\sin(ch)=ch$ and $\cos(ch)=1-\frac{c^{2}h^{2}}{2}.$ Now
it suffices to substitute $h$ and $\frac{h^{2}}{2}$ from these formulas
into (\ref{eq:_CircleOfCurvature}) to obtain the conclusion \[
\forall h\in D_{2}\pti\gamma(r+h)=\left(\gamma_{r}+\frac{\vec{n}_{r}}{c_{r}}\right)+\frac{1}{c_{r}}\cdot\left[\sin(c_{r}h)\vec{t}_{r}-\cos(c_{r}h)\vec{n}_{r}\right].\]
 In a similar way we can prove that any $f\in\Cc^{\infty}(\R,\R)$
can be written $\forall h\in D_{k}$ as \[
f(h)=\sum_{n=0}^{k}a_{n}\cdot\cos(nh)+\sum_{n=0}^{k}b_{n}\cdot\sin(nh),\]
so that now the idea of the Fourier series comes out in a natural
way.

\subsection{\noindent Commutation of differentiation and integration.}

\noindent This example derives from \citet{Koc,Lav}. Suppose we want
\emph{to discover} the derivative of the function \[
g(x):=\int_{\alpha(x)}^{\beta(x)}f(x,t)\diff{t}\qquad\forall x\in\R\]
 where $\alpha$, $\beta$ and $f$ are smooth functions. We can see
$g$ as a composition of smooth functions, hence we can apply the
derivation formula, i.e. Theorem \ref{thm:DerivationFormula}: \begin{align*}
g(x+h)= & \int_{\alpha(x+h)}^{\beta(x+h)}f(x+h,t)\diff{t}=\\
= & \int_{\alpha(x)+h\alpha'(x)}^{\alpha(x)}f(x,t)\diff{t}+h\cdot\int_{\alpha(x)+h\alpha'(x)}^{\alpha(x)}\frac{\partial f}{\partial x}(x,t)\diff{t}+\\
{} & +\int_{\alpha(x)}^{\beta(x)}f(x,t)\diff{t}+h\cdot\int_{\alpha(x)}^{\beta(x)}\frac{\partial f}{\partial x}(x,t)\diff{t}+\\
{} & +\int_{\beta(x)}^{\beta(x)+h\beta'(x)}f(x,t)\diff{t}+h\cdot\int_{\beta(x)}^{\beta(x)+h\beta'(x)}\frac{\partial f}{\partial x}(x,t)\diff{t}.\end{align*}
 Now we use $h^{2}=0$ to obtain e.g. (see Corollary \ref{cor:derivationFormulaForIntegrals_1D}):
\[
h\cdot\int_{\alpha(x)+h\alpha'(x)}^{\alpha(x)}\frac{\partial f}{\partial x}(x,t)\diff{t}=-h^{2}\cdot\alpha'(x)\cdot\frac{\partial f}{\partial x}(\alpha(x),t)=0\]
 and \[
\int_{\alpha(x)+h\alpha'(x)}^{\alpha(x)}f(x,t)\diff{t}=-h\cdot\alpha'(x)\cdot f(\alpha(x),t).\]
Calculating in an analogous way similar terms we finally obtain the
well known conclusion. Note that the final formula comes out by itself
so that we have \emph{{}``discovered''} it and not simply we have
proved it. From the point of view of artificial intelligence or from
the didactic point of view, surely this discovering is not a trivial
result.

\subsection{\noindent Schwarz's theorem.}

\noindent Using nilpotent infinitesimals we can obtain a simple and
meaningful proof of Schwarz's theorem. This simple example aims to
show how to manage some differences between our setting and SDG. Let
$f:V\freccia E$ be a $\Cc^{2}$ function between spaces of type $V=\R^{m}$,
$E=\R^{n}$ and $a\in V$, we want to prove that ${\rm d}^{2}{f}(a):V\times V\freccia E$
is symmetric. Take\begin{align*}
{} & k\in D_{2}\\
{} & h,j\in\D_{\infty}\\
{} & jkh\in D_{\ne0}\end{align*}
(e.g. we can take $k_{t}=\diff{t}_{2},h_{t}=j_{t}=\diff{t}_{4}$ so
that $jkh=\diff{t}$, see also Theorem \ref{thm:productOfPowers}).
Using $k\in D_{2}$, we have \begin{equation}
\begin{split}j\cdot f(x & +hu+kv)=\\
 & =j\cdot\left[f(x+hu)+k\,\partial_{v}f(x+hu)+\frac{k^{2}}{2}\partial_{v}^{2}f(x+hu)\right]\\
 & =j\cdot f(x+hu)+jk\cdot\partial_{v}f(x+hu)\end{split}
\label{eq: x+h+k}\end{equation}
where we used the fact that $k^{2}\in D$ and $j$ infinitesimal imply
$jk^{2}=0$. Now we consider that $jkh\in D$ so that any product
of type $jkhi$ is zero for every $i\in D_{\infty}$, so we obtain
\begin{equation}
jk\cdot\partial_{v}f(x+hu)=jk\cdot\partial_{v}f(x)+jkh\cdot\partial_{u}(\partial_{v}f)(x).\label{eq: jkh}\end{equation}
But $k\in D_{2}$ and $jk^{2}=0$ hence \[
j\cdot f(x+kv)-j\cdot f(x)=jk\cdot\partial_{v}f(x).\]
Substituting this in \eqref{eq: jkh} and hence in \eqref{eq: x+h+k}
we obtain \begin{equation}
\begin{split} & j\cdot\left[f(x+hu+kv)-f(x+hu)-f(x+kv)+f(x)\right]=\\
 & =jkh\cdot\partial_{u}(\partial_{v}f)(x).\end{split}
\label{eq: SecondOrderIncrementalRatio}\end{equation}
The left hand side of this equality is symmetric in $u,v$, hence
changing them we have \[
jkh\cdot\partial_{u}(\partial_{v}f)(x)=jkh\cdot\partial_{v}(\partial_{u}f)(x)\]
and thus we obtain the conclusion because $jkh\ne0$ and $\partial_{u}(\partial_{v}f)(x)$,
$\partial_{v}(\partial_{u}f)(x)\in E$. From (\ref{eq: SecondOrderIncrementalRatio})
it follows directly the classical limit relation \[
\lim_{t\to0^{+}}\frac{f(x+h_{t}u+k_{t}v)-f(x+h_{t}u)-f(x+k_{t}v)+f(x)}{h_{t}k_{t}}=\partial_{u}\partial_{v}f(x)\]

\subsection{Area of the circle and volumes of revolution.}

A more or less meaningful proof of the familiar formula for the area
of a circle depends on what axioms are assumed and how much general
the definitions are. In this example we want to show the possibility
to define suitable smooth functions using an infinitesimal property.
Let us assume the axioms for the real field $\R$; prove from them
the existence of the smooth functions $\sin$ and $\cos$; define
$\pi$ as a suitable zero of these functions (see e.g. \citet{Pro,Sil})
and define the length of an arc of circle of radius $r$, parametrized
by $x(\theta)=r\cdot\cos(\theta)$ and $y(\theta)=r\cdot\sin(\theta)$,
as the unique function $s$ that verifies\begin{align}
\left[s(\theta+k)-s(\theta)\right]^{2} & =\left[x(\theta+k)-x(\theta)\right]^{2}+\left[y(\theta+k)-y(\theta)\right]^{2}\quad\forall\theta\in\R\ \forall k\in D_{2}\label{eq:lengthOfACurve}\\
s(0) & =0.\label{eq:InitialCondForLength}\end{align}
This definition can be justified in the usual way using a (second
order!) infinitesimal right-angled triangle. The uniqueness of $s$
follows from \eqref{eq:lengthOfACurve} and \eqref{eq:InitialCondForLength},
the smoothness of $x$ and $y$, the second order infinitesimal Taylor's
formula and the cancellation law (Theorem \ref{thm:firstCancellationLaw}):\[
k^{2}\cdot\dot{s}(\theta)=\dot{x}(\theta)\cdot k^{2}+\dot{y}(\theta)\cdot k^{2}\quad\forall k\in D_{2}.\]
From this and \eqref{eq:InitialCondForLength} we obtain the usual
formula for $s$ that, in our particular case, gives $s(\theta)=r\cdot\theta$.
Now we can think the area $A(\theta+h)-A(\theta)$ of a first order
infinitesimal sector of the circle as the area of the isosceles triangle
with sides of length $r$ and base $s(\theta+h)-s(\theta)$. In fact,
if $P(\theta)=\left(r\sin\theta,r\cos\theta\right)$, then $P(\theta+h)=P(\theta)+h\cdot\vec{t}(\theta)$,
where $\vec{t}$ is the tangent vector, so that in $[\theta,\theta+h]$,
$h\in D$, the circle is made of linear segments. Therefore, the area
$A(\theta)$ can be defined as the unique function that verifies\begin{align*}
A(\theta+h)-A(\theta) & =\frac{1}{2}\left[s(\theta+h)-s(\theta)\right]\cdot r\cos\left(\frac{h}{2}\right)\quad\forall\theta\in\R\ \forall h\in D\\
A(0) & =0.\end{align*}
From this and the derivation formula we get\begin{align*}
h\cdot A'(\theta) & =\frac{1}{2}hr\cdot s'(\theta)\\
A(\theta) & =\frac{1}{2}\int_{0}^{\theta}r\cdot s(u)\diff{u}.\end{align*}
In our case we get $A(\theta)=\frac{1}{2}r^{2}\cdot\theta$ and hence
the searched formula for $\theta=2\pi$.\\
Analogously we can prove the familiar formula for volumes of revolution
of a parametrized curve $\gamma(u)=\left(x(u),y(u)\right)$, $u\in[a,b]$,
around the $x$-axis. Let us define the volume as the unique smooth
function $V$ that verifies\begin{align}
V(u+h)-V(u) & =h\cdot\pi\cdot y(u)^{2}+\frac{1}{2}\left[h\cdot\pi\cdot y(u+h)^{2}-h\cdot\pi\cdot y(u)^{2}\right]\label{eq:volumeOfRevolutionDef}\\
V(0) & =0\label{eq:volumeOfRevolutionInitialCondition}\end{align}
for every $u\in[a,b]$ and $h\in D$. This definition can be intuitively
justified saying that the volume of the sector of revolution between
$u$ and $u+h$ can be calculated as the sum of the cylinder of radius
$y(u)$ and height $h$ plus one half of the difference between the
cylinder of radius $y(u+h)$ and height $h$ and that of radius radius
$y(u)$ and the same height. Implicitly, we are using the straightness
of the curve $\gamma$ in $[u,u+h]$. From \eqref{eq:volumeOfRevolutionDef}
and the property $h^{2}=0$ we easily obtain that $V'(u)=\pi\cdot y(u)^{2}$
and hence the usual formula using \eqref{eq:volumeOfRevolutionInitialCondition}.

\subsection{Curvature.}

Let us consider the usual smooth parametrized curve $\gamma(u)=\left(x(u),y(u)\right)$
for $u\in[a,b]$. Let $\phi(u)\in[0,\pi]$ be the non-oriented angle
(i.e. the one defined by the scalar product) between the tangent vector
$\vec{t}=(\dot{x},\dot{y})$ and the unit vector $\vec{i}$ of the
$x$-axis, so that\[
\sqrt{\dot{x}^{2}+\dot{y}^{2}}\cdot\cos\phi=\dot{x}.\]
Multiplying this equality by $\sin\phi$ we easily obtain\begin{equation}
\dot{y}\cdot\cos\phi=\dot{x}\cdot\sin\phi.\label{eq:fundamentalRelationOfPhi}\end{equation}
It is well known that the curvature of $\gamma$ at the point $u\in[a,b]$
can be calculated as the rate of change of the non-oriented angle
$\phi(u)$ with respect to an infinitesimal variation in arc length
$s(u)$ defined by the analogous of \eqref{eq:lengthOfACurve} and
\eqref{eq:InitialCondForLength}. These {}``rate of changes'' can
be defined in $\ER$ as the unique (if it exists) standard $c(u)\in\R$
defined by\[
c(u)\cdot\left[s(u+h)-s(u)\right]=\phi(u+h)-\phi(u)\quad\forall h\in D.\]
Indeed, from the cancellation law, i.e. Theorem \ref{thm:firstCancellationLaw},
there exists at most one such $c(u)\in\R$ verifying this property.
Because of this uniqueness we can also use the notation\begin{equation}
c(u)=\frac{\phi(u+h)-\phi(u)}{s(u+h)-s(u)}.\label{eq:rateOfChange}\end{equation}
These ratios generalize the usual ratios between real numbers (see
\citet{Gio4} for more details). From \eqref{eq:rateOfChange} and
the derivation formula we get $c(u)=\frac{h\cdot\phi'(u)}{h\cdot s'(u)}=\frac{\phi'(u)}{s'(u)}$
whatever $h\in D_{\ne0}$ we choose. From this and the relation \eqref{eq:fundamentalRelationOfPhi}
(without using infinitesimals, but using standard differential calculus)
we can obtain the usual formula $c=\frac{\dot{x}\ddot{y}-\dot{y}\ddot{x}}{\left(\dot{x}^{2}+\dot{y}^{2}\right)^{3/2}}$
at each point $u\in[a,b]$ where $\phi(u)\ne\frac{\pi}{2}$ and $\dot{\gamma}(u)\ne\underline{0}$.

\subsection{Stretching of a spring (and center of pressure).}

If $f:[a,b]\freccia\R$ is a smooth function and we define $J(x):=\int_{0}^{x}f(s)\diff{s}$,
then Corollary \ref{cor:derivationFormulaForIntegrals_1D} and a trivial
calculation with the derivation formula give\begin{equation}
J(x+h)-J(x)=\frac{1}{2}\left[f(x+h)+f(x)\right]\quad\forall h\in D.\label{eq:averageValue}\end{equation}
The right-hand side of \eqref{eq:averageValue} is interpreted as
the average value of $f$ in the infinitesimal interval $[x,x+h]$.
Analogous equalities can be obtain in the $d$-dimensional case using
suitable generalizations of the above cited corollary: e.g. if $d=2$
we have to use\[
\int_{0}^{h}\int_{0}^{k}f(x,y)\diff{x}\diff{y}=hk\cdot f(0,0)\quad\forall h,k\in D_{\infty}:\ h\cdot k\in D.\]
These equalities are used by \citet{Bel} to calculate the center
of pressure of a plane area and the work done in stretching a spring.
The meaningfulness of such examples is however doubtful because they
can be summarized saying: assume to have a smooth $J$ satisfying
\eqref{eq:averageValue}; deduce from this and from the assumption
$J(0)=0$ that $J'(x)=f(x)$. There is no real use of infinitesimals
in this type of reasoning in every case where the definition $J(x):=\int_{0}^{x}f(s)\diff{s}$
is customary, like in the cited examples.

\subsection{The wave equation.}

The deduction of the wave equation in the framework of Fermat reals
is very interesting for two main reasons. Firstly, in the classical
deduction (see e.g. \citet{Vla}) there are some approximations tied
with Hook's law. Is it possible to make them rigorous using $\ER$?
Do we gain something using this increased rigour? E.g.: how can we
formalize the approximated equalities used in the classical deduction?
In what a sense is the wave equation an approximated equality valid
for small oscillations only?

Secondly, at the end of our deduction we will stress the physical
principles as important mathematical assumptions of a suitable theorem.
We are hence naturally taken to ask if these natural assumptions (some
of which formulated using the infinitesimals of $\ER$) really have
a model. In this way, we will see that no standard smooth function
can satisfy these hypothesis, but we are forced to consider a non-standard
one. E.g. $f(x)=h\cdot\sin(x)$ for $x\in\ER$ and $h\in D_{\infty}$
is an example of a non-standard smooth function; let us note that
it is obtained by the standard smooth function $g(y,x):=y\cdot\sin(x)$,
$x$, $y\in\R$, by extension to $\ER^{2}$ and fixing one of its
variables to a non-standard parameter $h\in D_{\infty}$:\[
f(x)=\ext{g}(h,x)\quad\forall x\in\ER.\]
This will motivate strongly the further development of the theory
of Fermat reals, in the direction of a more general theory including
also these new smooth non-standard functions.

Let us start considering a string making small transversal oscillations
around its equilibrium position located on the interval $[a,b]$ of
the $x$ axis, for $a$, $b\in\R$, $a<b$. By hypotheses, string's
position $s_{t}\subseteq\ER^{2}$ is always represented by the graph
of a given curve $\gamma:[a,b]\times[0,+\infty)\freccia\ER^{2}$ (where
$[a,b]=\left\{ x\in\ER\,|\, a\le x\le b\right\} $ and $[0,+\infty)=\left\{ x\in\ER\,|\,0\le x\right\} $;
in the following, we will always use these notations for intervals
to identify the corresponding subsets of $\ER$, and not of $\R$,
and we will also use the notation $\gamma_{xt}:=\gamma(x,t)$):\[
s_{t}=\left\{ \gamma_{xt}\in\ER^{2}\,|\, a\le x\le b\right\} \quad\forall t\in[0,+\infty).\]
Moreover, the curve $\gamma$ is supposed to be injective with respect
to the parameter $x\in(a,b)$:\[
\gamma_{x_{1}t}\ne\gamma_{x_{2}t}\quad\forall t\in[0,+\infty)\ \forall x_{1},x_{2}\in(a,b):\ x_{1}\ne x_{2},\]
so that the order relation on $(a,b)$ implies an order relation on
the support $s_{t}$. For every pair of points $p=\gamma_{x_{p}t}$,
$q=\gamma_{x_{q}t}\in s_{t}$ on the string at time $t$, we can define
the sub-bodies:\begin{align*}
\overrightarrow{p} & :=\left\{ \gamma_{xt}\,|\, x_{p}\le x\le b\right\} \\
\overleftarrow{p} & :=\left\{ \gamma_{xt}\,|\, a\le x\le x_{p}\right\} \\
\overrightarrow{pq} & :=\left\{ \gamma_{xt}\,|\, x_{p}\le x\le x_{q}\right\} \end{align*}
corresponding respectively to the parts of the string that follows
the point $p\in s_{t}$, that precedes the same point and that lies
between the point $p\in s_{t}$ and the point $q\in s_{t}$. It is
usually implicitly clear that e.g. every sub-body of the form $\overrightarrow{p}$
exerts a force on each sub-body with which it is in contact, i.e.
of the form $\overrightarrow{pq}$ or $\overleftarrow{p}$. Moreover,
the force $\mathbf{F}(A,B)\in\ER^{2}$ that the sub-body $A$ exerts
on the sub-body $B$ verifies the following equalities (see e.g. \citet{Tru}):\begin{align}
\mathbf{F}(\overrightarrow{pq},\overleftarrow{p}) & =\mathbf{F}(\overrightarrow{p},\overleftarrow{p})\label{eq:firstForceLaw}\\
\mathbf{F}(\overrightarrow{q},\overrightarrow{pq}) & =\mathbf{F}(\overrightarrow{q},\overleftarrow{q})\label{eq:secondForceLaw}\\
\mathbf{F}(\overleftarrow{p},\overrightarrow{pq}) & =-\mathbf{F}(\overrightarrow{pq},\overleftarrow{p})\quad\text{(action-reaction principle)}\label{eq:action-reaction}\end{align}
for every pair of points $p$, $q\in s_{t}$ and every time $t\in[0,+\infty)$.
Using this formalism, the tension at the point $\gamma_{xt}\in s_{t}$
at time $t\in[0,+\infty)$ can now be defined in the following way\begin{equation}
\mathbf{T}(x,t):=\mathbf{F}(\overrightarrow{\gamma_{xt}},\overleftarrow{\gamma_{xt}}).\label{eq:defOfTension}\end{equation}
Now, let us consider the infinitesimal sub-body $\overrightarrow{x,x+\delta x}:=\overrightarrow{\gamma_{xt}\gamma_{x+\delta x,t}}\subseteq s_{t}$
located at time $t$ between the points $\gamma_{xt}\in s_{t}$ and
$\gamma_{x+\delta x,t}\in s_{t}$, where $\delta x\in D$ is a generic
first order infinitesimal. On this infinitesimal sub-body, mass forces
of linear density $\mathbf{G}:[a,b]\times[0,+\infty)\freccia\ER^{2}$
act, so that Newton's law can be written as\begin{equation}
\rho\cdot\delta x\cdot\frac{\partial^{2}\gamma}{\partial t^{2}}=\mathbf{F}(\overleftarrow{\gamma_{xt}},\overrightarrow{x,x+\delta x})+\mathbf{F}(\overrightarrow{\gamma_{x+\delta x,t}},\overrightarrow{x,x+\delta x})+\mathbf{G}\cdot\rho\cdot\delta x,\label{eq:NewtonLaw}\end{equation}
where $\rho:[a,b]\times[0,+\infty)\freccia\ER$ is the linear mass
density and where, if not otherwise indicated, all the functions are
calculated at $(x,t)\in(a,b)\times[0,+\infty)$. Of course, the contact
forces appearing in Newton's law are due to the interaction of the
infinitesimal sub-body with other sub-bodies in contact with its border\[
\partial\left[\overrightarrow{x,x+\delta x}\right]=\left\{ \gamma_{xt},\gamma_{x+\delta x,t}\right\} \subseteq\ER^{2}.\]
Using action-reaction principle \eqref{eq:action-reaction} and the
equality \eqref{eq:secondForceLaw}, with $q=\gamma_{x+\delta x,t}$
and $p=\gamma_{xt}$ so that $\overrightarrow{pq}=\overrightarrow{x,x+\delta x}$,
from \eqref{eq:NewtonLaw} we have\[
\rho\cdot\delta x\cdot\frac{\partial^{2}\gamma}{\partial t^{2}}=-\mathbf{F}(\overrightarrow{x,x+\delta x},\overleftarrow{\gamma_{xt}})+\mathbf{F}(\overrightarrow{\gamma_{x+\delta x,t}},\overleftarrow{\gamma_{x+\delta x,t}})+\mathbf{G}\cdot\rho\cdot\delta x.\]
Using \eqref{eq:firstForceLaw} and the definition \eqref{eq:defOfTension}
of tension we get\begin{align}
\rho\cdot\delta x\cdot\frac{\partial^{2}\gamma}{\partial t^{2}} & =-\mathbf{F}(\overrightarrow{\gamma_{xt}},\overleftarrow{\gamma_{xt}})+\mathbf{F}(\overrightarrow{\gamma_{x+\delta x,t}},\overleftarrow{\gamma_{x+\delta x,t}})+\mathbf{G}\cdot\rho\cdot\delta x\nonumber \\
 & =-\mathbf{T}(x,t)+\mathbf{T}(x+\delta x,t)+\mathbf{G}\cdot\rho\cdot\delta x.\label{eq:NewtonWithTension}\end{align}
Up to this point of the deduction we have not used neither the hypotheses
of small oscillations nor that of transversal oscillations. The second
one can be easily introduced with the hypotheses\begin{equation}
\mathbf{G}(x,t)\cdot\vec{e}_{1}=0\quad\forall x,t,\label{eq:transversalOscillations}\end{equation}
where $(\vec{e}_{1},\vec{e}_{2})$ are the axis unit vectors. Using
the notation $\phi(x,t)$ for the non-oriented angle between the tangent
unit vector $\mathbf{t}(x,t)$ at the point $\gamma_{xt}$ and the
$x$ axes (see \eqref{eq:fundamentalRelationOfPhi}), the hypotheses
of small oscillations can be formalized with the assumption\begin{equation}
\phi(x,t)\in D\quad\forall x,t.\label{eq:smallOscillations}\end{equation}
This will permit to reproduce the classical deduction in the most
faithful way (even if, as we will see later, a weaker assumption can
be considered). Moreover, in the classical deduction of the wave equation,
one considers only curves of the form $\gamma_{xt}=(x,u(x,t))$. In
this way from \eqref{eq:fundamentalRelationOfPhi} and the derivation
formula we have\begin{align*}
\frac{\partial\gamma_{2}}{\partial x}\cdot\cos\phi & =\sin\phi\\
\frac{\partial\gamma_{2}}{\partial x} & =\phi\in D\end{align*}
so that $\left(\frac{\partial\gamma_{2}}{\partial x}\right)^{2}=0$
and hence the total length of the string becomes:\begin{equation}
L=\int_{a}^{b}\sqrt{1+\left[\frac{\partial\gamma_{2}}{\partial x}(x,t)\right]^{2}}\diff{x}=b-a\quad\forall t\in[0,+\infty).\label{eq:lengthForSmallOscillations}\end{equation}
By Hook's law, this justifies that the tension can be assumed to have
a constant modulus $T$, not depending neither by the position $x$
nor by the time $t$:\begin{equation}
\mathbf{T}(x,t)=T\cdot\mathbf{t}(x,t)\quad\forall x\in(a,b)\ \forall t\in[0,+\infty).\label{eq:hypOfConstantModulusForTension}\end{equation}
A tension $\mathbf{T}$ parallel to the tangent vector is the second
part of the hypothesis about non transversal oscillations of the string.
Let us note explicitly that the only standard continuous function
verifying the equality $L=b-a$ is the constant one, so the function
$u:[a,b]\times[0,+\infty)\freccia\ER$ has to be understood as a non-standard
one; later we will do further considerations about this important
point. Projecting the equation \eqref{eq:NewtonWithTension} on the
$y$ axis, we obtain\begin{align*}
\rho\cdot\delta x\cdot\frac{\partial^{2}u}{\partial t^{2}} & =-T\cdot\mathbf{t}(x,t)\cdot\vec{e}_{2}+T\cdot\mathbf{t}(x+\delta x,t)\cdot\vec{e}_{2}+\mathbf{G}\cdot\vec{e}_{2}\cdot\rho\cdot\delta x\\
 & =-T\sin\phi(x,t)+T\cdot\sin\phi(x+\delta x,t)+G\cdot\rho\cdot\delta x.\end{align*}
But $\sin\phi=\phi=\frac{\partial u}{\partial x}$ because $\phi\in D$
is a first order infinitesimal, hence\begin{align}
\rho\cdot\delta x\cdot\frac{\partial^{2}u}{\partial t^{2}} & =T\cdot\left[\frac{\partial u}{\partial x}(x+\delta x,t)-\frac{\partial u}{\partial x}(x,t)\right]+G\cdot\rho\cdot\delta x\nonumber \\
 & =\left[T\cdot\frac{\partial^{2}u}{\partial x^{2}}(x,t)+G\cdot\rho\right]\cdot\delta x.\label{eq:finalBeforeCancelling}\end{align}
We cannot use the cancellation law with $\delta x\in D$ to obtain
the final result, because, as we mentioned above, the function $u(x,t)\in\ER$
can assume non standard values, so it is time to clarify some points.
As mentioned above, there does not exist a standard smooth function
verifying all the assumptions or the physical principles we have used.
Of course, everything depends by how we formalize the classical informal
deduction used in elementary physics: e.g. we have chosen to use an
equality sign in \eqref{eq:lengthForSmallOscillations} instead of
an approximated equality; anyway we have to consider that if we use
$\simeq$ to write \eqref{eq:lengthForSmallOscillations}, then the
problem becomes how to make more precise, physically, numerically
or mathematically, this approximation; moreover, if we use an approximation
sign in \eqref{eq:lengthForSmallOscillations}, then we consistently
must use the same sign both in \eqref{eq:hypOfConstantModulusForTension}
and therefore in the final wave equation. Nevertheless, smooth non
standard functions can verify all the hypothesis and physical principles
we have considered: e.g. the function $u(x,t):=u_{0}\sin(x+\omega\cdot t)$
is one of these if the maximum amplitude $u_{0}\in D$ and if $\rho$
is constant, $G=0$ and $T=\omega^{2}\rho$.
\begin{defn}
If $X\subseteq\ext{\R^{\sf x}}$ and $Y\subseteq\ext{\R^{\sf y}}$
then we say that \[
f:X\xfrecciad{}Y\text{ is (non standard) smooth}\]
 iff $f$ maps $X$ in $Y$ and for every $x_{0}\in X$ we can write
\begin{equation}
f(x)=\ext{g}\langle p,x\rangle\quad\forall x\in\ext{V}\cap X\label{eq:localFormForArrowsOfSERn}\end{equation}
 for some \begin{align*}
{} & V\text{ open in }\R^{\sf x}\text{ such that }x_{0}\in\ext{V}\\
{} & p\in\ext{U},\text{ where }U\text{ is open in }\R^{\sf p}\\
{} & g\in\Cc^{\infty}(U\times V,\R^{\sf y}),\end{align*}
where $\langle-,-\rangle:([x]_{\sim},[y]_{\sim})\in\ext{U}\times\ext{V}\longmapsto[(x,y)]_{\sim}\in\ext{(U\times V)}$
(see Definition \ref{def:equalityInFermatReals} for the relation
$\sim$).
\end{defn}
In other words locally\emph{ }a smooth function $f:X\freccia Y$ from
$X\subseteq\ext{\R^{\sf x}}$ to $Y\subseteq\ext{\R^{\sf y}}$ is
constructed in the following way:
\begin{enumerate}
\item start with an ordinary standard function $g\in\Cc^{\infty}(U\times V,\R^{\sf y})$,
with $U$ open in $\R^{\sf p}$ and $V$ open in $\R^{\sf x}$. The
space $\R^{\sf p}$ has to be thought as a space of parameters for
the function $g$;
\item consider its Fermat extension obtaining $\ext{g}:\ext{(U\times V)}\freccia\ext{\R^{\sf y}}$;
\item consider the composition $\ext{g}\circ\langle-,-\rangle:\ext{U}\times\ext{V}\freccia\ext{\R^{\sf y}}$,
where $\langle-,-\rangle$ is the isomorphism $\ext{U}\times\ext{V}\simeq\ext{(U\times V)}$
defined by $\langle[x]_{\sim},[y]_{\sim}\rangle=[(x,y)]_{\sim}$;
we will always use the identification $\ext{U}\times\ext{V}=\ext{(U\times V)}$,
so we will write simply $\ext{g}(p,x)$ instead of $\ext{g}\langle p,x\rangle$.
\item fix a parameter $p\in\ext{U}$ as a first variable of the previous
composition, i.e. consider $\ext{g}\langle p,-\rangle:\ext{V}\freccia\ext{\R^{\sf y}}$.
Locally, the map $f$ is of this form: $f=\ext{g}\langle p,-\rangle=\ext{g}(p,-)$.
\end{enumerate}
Because $p=\st{p}+h$, with $h\in D_{\infty}$, applying the infinitesimal
Taylor's formula to variable $p$ for the function $\ext{g(p,x)}$
it is not hard to prove the following Theorem, that clarifies further
the form of these non standard smooth functions, because it states
that they can be seen locally as {}``infinitesimal polynomials with
smooth coefficients'':
\begin{thm}
\label{thm:infPolynomialsWithSmoothCoefficients}Let $X\subseteq\ER^{\sf x}$
and $f:X\freccia\ER^{n}$ a map. Then it results that\[
f:X\freccia\ER^{n}\text{ is non standard smooth}\]
 if and only if for every $x_{0}\in X$ we can write\begin{equation}
f(x)=\sum_{\substack{|q|\le k\\
q\in\N^{d}}
}a_{q}(x)\cdot p^{q}\quad\forall x\in\ext{V}\cap X,\label{eq:thesisInfPolyWithSmoothCoeff}\end{equation}
for suitable:
\begin{enumerate}
\item $d$, $k\in\N$
\item $p\in D_{k}^{d}$
\item $V$ open subset of $\R^{\sf x}$ such that $x_{0}\in\ext{V}$
\item $(a_{q})_{\substack{|q|\le k\\
q\in\N^{d}}
}$ family of $\mathcal{C}^{\infty}(V,\R^{n})$.
\end{enumerate}
\end{thm}
In other words, every smooth function $f:X\freccia\ER^{n}$ can be
constructed locally starting from some {}``infinitesimal parameters''\[
p_{1},\ldots,p_{d}\in D_{k}\]
and from ordinary smooth functions\[
a_{q}\in\mathcal{C}^{\infty}(V,\R^{n})\]
and using polynomial operation only with $p_{1}$, ..., $p_{d}$ and
with coefficients $a_{q}(-)$. Roughly speaking, we can say that they
are {}``infinitesimal polynomials with smooth coefficients. The polynomials
variables act as parameters only''.

As it is natural to expect, several notions of differential and integral
calculus, including their infinitesimal versions, can be extended
to this type of new smooth function (for more details, see the preprint
\citet{Gio4}), and these results will be presented in future works.
In this sense, this deduction of the wave equation motivates strongly
the future development of the theory of Fermat reals.

On the other hand, we have to understand what type of cancellation
law we can apply to \eqref{eq:finalBeforeCancelling}. For this end,
we have to define the notion of equality up to $k$-th order infinitesimals:
\begin{defn}
\label{def:upTo_k-thOrderInfinitesimals} Let $m=\st{m}+\sum_{i=1}^{N}\st{m_{i}}\cdot\diff{t}_{\omega_{i}(m)}$
be the decomposition of $m\in\ER$ and $k\in\R_{\ge0}\cup\{\infty\}$,
then\[
{\displaystyle \iota_{k}m:=\iota_{k}(m):=\st{m}+\sum_{\substack{i=1\\
\omega_{i}(m)>k}
}^{N}\st{m_{i}}\cdot\diff{t_{\omega_{i}(m)}}}.\]

Finally if $x$, $y\in\ER$, we will say $x=_{k}y$ iff $\iota_{k}x=\iota_{k}y$
in $\ER$, and we will read it as \emph{$x$ is equal to $y$ up to
$k$-th order infinitesimals.}
\end{defn}
In other words, as it is easy to prove, we have\[
x=_{k}y\quad\iff\quad\st{x}=\st{y}\ \text{ and }\ \omega(x-y)\le k.\]
Therefore, if we denote with\[
I_{k}:=\left\{ x\in D_{\infty}\,|\,\omega(x)\le k\right\} ,\]
the set of all the infinitesimal of order less that or equal $k$
(let us note that $I_{k}\subset D_{k}$), then we have that $x=_{k}y$
if and only if $x-y\in I_{k}$. Equality up to $k$-th order infinitesimal
is of course an equivalence relation and preserves all the ring operations
of $\ER$. More in general these equalities are preserved by smooth
functions $f:\ER\freccia\ER$:\[
x=_{k}y\then f(x)=_{k}f(y).\]
Using this notion, it is not hard to prove the following cancellation
law up to $k$-th order infinitesimals.
\begin{thm}
\label{thm:generalCancellationLaw}Let $m\in\ER$, $n\in\N_{>0}$,
$j\in\N^{n}\setminus\{\underline{0}\}$ and $\alpha\in\R_{>0}^{n}$.
Moreover let us consider $k\in\R$ defined by \begin{equation}
\frac{1}{k}+\sum_{i=1}^{n}\frac{j_{i}}{\alpha_{i}+1}=1\label{eq:conditionToFind_k}\end{equation}

\noindent then
\begin{enumerate}
\item \label{enu:generalThm_hTojAndProductWithIota}$\forall h\in D_{\alpha_{1}}\times\cdots\times D_{\alpha_{n}}\pti h^{j}\cdot m=h^{j}\cdot\iota_{k}m$ 
\item \textup{\emph{\label{enu:generalCancellationLaw}If $h^{j}\cdot m=0$
for every $h\in D_{\alpha_{1}}\times\cdots\times D_{\alpha_{n}}$,
then $m=_{k}0$}}
\end{enumerate}
\end{thm}
\noindent E.g. if $n=1$ and $\alpha_{1}=j_{1}=1$ we have $k=2$
and hence\[
\forall h\in D\pti h\cdot m=h\cdot\iota_{2}m\]
\begin{equation}
\left(\forall h\in D\pti h\cdot m=0\right)\iff m=_{2}0.\label{eq:cancellationLawUpTo2}\end{equation}
Using \eqref{eq:cancellationLawUpTo2} in \eqref{eq:finalBeforeCancelling}
we obtain the final conclusion\begin{equation}
\rho\cdot\frac{\partial^{2}u}{\partial t^{2}}=_{2}T\cdot\frac{\partial^{2}u}{\partial x^{2}}+G\cdot\rho\quad\forall x\in(a,b)\ \forall t\in(0,+\infty).\label{eq:waveEquationsUpTo2}\end{equation}

It is also interesting to note that not only small oscillations of
the string implies \eqref{eq:waveEquationsUpTo2}, but the converse
is also true: the equation \eqref{eq:waveEquationsUpTo2} implies
that necessary we must have small oscillations of the string, i.e.
that $\phi(x,t)\in D_{\infty}$. Moreover, using the equality $=_{2}$
up to second order infinitesimals, all the classical approximation
tied with Hook's law, now become more clear. Indeed, we have the following
\begin{thm}
\label{thm:waveEquationTheorem}Let $a$, $b\in\R$, with $a<b$;
let $\gamma:[a,b]\times[0,+\infty)\freccia\ER^{2}$, $\rho:[a,b]\times[0,+\infty)\freccia\ER$
and $\mathbf{G},\mathbf{T}:[a,b]\times[0,+\infty)\freccia\ER^{2}$
be non standard smooth functions and $T\in\ER$ be an invertible Fermat
real. Let us suppose that the first component $\gamma_{1}$ of the
curve is of the form\begin{equation}
\gamma_{1}(x,t)=\left[1+\alpha(t)\right]\cdot x+\beta(t)\quad\forall x,t,\label{eq:hypInfPerturbationOfIdentity}\end{equation}
with $\alpha(t)\in I_{2}$. Then the unit tangent vector $\mathbf{t}(x,t)$
to the curve $\gamma$ exists and we can further suppose that the
relations\begin{align}
\mathbf{T}(x,t) & =_{2}T\cdot\mathbf{t}(x,t)\label{eq:tensionInTheorem}\\
\rho\cdot\delta x\cdot\frac{\partial^{2}\gamma_{xt}}{\partial t^{2}} & =\mathbf{T}(x+\delta x,t)-\mathbf{T}(x,t)+\mathbf{G}\cdot\rho\cdot\delta x,\label{eq:NewtonIntheorem}\end{align}
holds for a every point $(x,t)\in(a,b)\times[0,+\infty)$ and for
every $\delta x\in D$. Finally, let us suppose that\[
\frac{\partial\phi}{\partial x}(x,t)\text{ is invertible.}\]
Then at this point $(x,t)$ the following sentences are equivalent
\begin{enumerate}
\item \label{enu:waveEqInTheorem}$\rho(x,t)\cdot\frac{\partial^{2}\gamma_{2}}{\partial t^{2}}(x,t)=_{2}T\cdot\frac{\partial^{2}\gamma_{2}}{\partial x^{2}}(x,t)+G_{2}(x,t)\cdot\rho(x,t)$
\item \label{enu:phiMustBeInfinitesimal}$\phi(x,t)\in I_{4}$.
\end{enumerate}
Finally, if \eqref{enu:phiMustBeInfinitesimal} holds for every $(x,t)\in(a,b)\times[0,+\infty)$,
then\begin{align*}
\text{\emph{length}}(\gamma_{-,t}) & =_{2}b-a.\end{align*}

\end{thm}
\noindent To simplify the proof of this result, we need two lemmas.
\begin{lem}
\label{lem:equality2ImpliesEqualityOfDifferentials}Let $a$, $b\in\R$
with $a<b$ and let $f$, $g:(a,b)\freccia\ER$ be non standard smooth
functions such that\[
f(x)=_{2}g(x)\quad\forall x\in(a,b).\]
Then\[
f(x+h)-f(x)=g(x+h)-g(x)\quad\forall h\in D\ \forall x\in(a,b)\]

\end{lem}
\medskip{}

\begin{lem}
\label{lem:m_cos3Equal_m_up2isEquivalentToOrder4}Let $m$, $h\in\ER$,
and suppose that $m$ is invertible and $0\le h\le\pi$, then the
following properties are equivalent:
\begin{enumerate}
\item $m\cdot\cos^{3}h=_{2}m$
\item $h\in I_{4}$.
\end{enumerate}
\end{lem}
\noindent \textbf{Proof of Theorem \ref{thm:waveEquationTheorem}:}
We firstly note that, assuming \eqref{eq:hypInfPerturbationOfIdentity},
the tangent vector $\mathbf{t}(x,t)$ always exists in $\ER$. In
fact we have $\frac{\partial\gamma_{1}}{\partial x}(x,t)=1+\alpha(t)$
so that both $\frac{\partial\gamma_{1}}{\partial x}(x,t)$ and $\left[\frac{\partial\gamma_{1}}{\partial x}(x,t)\right]^{2}+\left[\frac{\partial\gamma_{2}}{\partial x}(x,t)\right]^{2}$
are invertible; we can hence take its square root and then the inverse
to define the unit tangent vector. Now we prove that \eqref{enu:waveEqInTheorem}
implies \eqref{enu:phiMustBeInfinitesimal}. Let us take a generic
$\delta x\in D$. Projecting \eqref{eq:NewtonIntheorem} on $\vec{e}_{2}$
we get\[
\rho\cdot\delta x\cdot\frac{\partial^{2}\gamma_{2}}{\partial t^{2}}=\mathbf{T}(x+\delta x,t)\cdot\vec{e}_{2}-\mathbf{T}(x,t)\cdot\vec{e}_{2}+G_{2}\cdot\rho\cdot\delta x.\]
But from \eqref{eq:tensionInTheorem} and because smooth operations
preserve $=_{2}$, we get $\mathbf{T}\cdot\vec{e}_{2}=_{2}T\cdot\mathbf{t}\cdot\vec{e}_{2}$.
Therefore, from Lemma \ref{lem:equality2ImpliesEqualityOfDifferentials}
we obtain \begin{align*}
\mathbf{T}(x+\delta x,t)\cdot\vec{e}_{2}-\mathbf{T}(x,t)\cdot\vec{e}_{2} & =T\cdot\mathbf{t}(x+\delta x,t)\cdot\vec{e}_{2}-T\cdot\mathbf{t}(x,t)\cdot\vec{e}_{2}\\
 & =T\cdot\sin\phi(x+\delta x,t)-T\cdot\sin\phi(x,t)\end{align*}
\begin{equation}
\rho\cdot\delta x\cdot\frac{\partial^{2}\gamma_{2}}{\partial t^{2}}=T\cdot\sin\phi(x+\delta x,t)-T\cdot\sin\phi(x,t)+G_{2}\cdot\rho\cdot\delta x.\label{eq:1FromNewtonInTheorem}\end{equation}
On the other hand, we can multiply \eqref{enu:waveEqInTheorem} by
$\delta x$ (so that $=_{2}$ becomes $=$, see Theorem \ref{thm:generalCancellationLaw})
obtaining\begin{align}
\rho\cdot\delta x\cdot\frac{\partial^{2}\gamma_{2}}{\partial t^{2}} & =T\cdot\left[\frac{\partial\gamma_{2}}{\partial x}(x+\delta x,t)-\frac{\partial\gamma_{2}}{\partial x}(x,t)\right]+G_{2}\cdot\rho\cdot\delta x\nonumber \\
 & =T\cdot\tan\phi(x+\delta x,t)\cdot\frac{\partial\gamma_{1}}{\partial x}(x+\delta x,t)-T\tan\phi(x,t)\cdot\frac{\partial\gamma_{1}}{\partial x}(x,t)+G_{2}\cdot\rho\cdot\delta x,\label{eq:2FromWaveInTheorem}\end{align}
Equating \eqref{eq:1FromNewtonInTheorem} and \eqref{eq:2FromWaveInTheorem}
and canceling $T$ we get\[
\sin\phi(x+\delta x,t)-\sin\phi(x,t)=\tan\phi(x+\delta x,t)\cdot\frac{\partial\gamma_{1}}{\partial x}(x+\delta x,t)-\tan\phi(x,t)\cdot\frac{\partial\gamma_{1}}{\partial x}(x,t)\]
\begin{align}
\delta x\cdot\cos\phi\cdot\frac{\partial\phi}{\partial x} & =\delta x\cdot\frac{1}{\cos^{2}\phi}\cdot\frac{\partial\phi}{\partial x}\cdot\frac{\partial\gamma_{1}}{\partial x}(x,t)+\tan\phi\cdot\frac{\partial^{2}\gamma_{1}}{\partial x^{2}}(x,t)\nonumber \\
 & =\delta x\cdot\frac{1}{\cos^{2}\phi}\cdot\frac{\partial\phi}{\partial x}\cdot\left[1+\alpha(t)\right]\nonumber \\
 & =\delta x\cdot\frac{1}{\cos^{2}\phi}\cdot\frac{\partial\phi}{\partial x}\label{eq:cosAnd1OverCosSquare}\end{align}
where, as usual, every function, if not otherwise indicated, is calculated
at $(x,t)$. Let us note that, in \eqref{eq:cosAnd1OverCosSquare}
we have used the property $\delta x\cdot\alpha(t)=0$ because $\delta x\in D$
and $\alpha(t)\in I_{2}$; moreover, from \eqref{eq:fundamentalRelationOfPhi}
if $\phi=\frac{\pi}{2}$ we would have $\frac{\partial\gamma_{2}}{\partial x}\cdot\cos\phi=0=\frac{\partial\gamma_{1}}{\partial x}\cdot\sin\phi=1+\alpha(t)$,
which is impossible because $\alpha(t)\in D_{\infty}$. Setting, for
simplicity, $m:=\frac{\partial\phi}{\partial x}(x,t)\in\ER$, from
\eqref{eq:cosAnd1OverCosSquare} and canceling $\delta x$, we have\begin{equation}
m\cdot\cos^{3}\phi=_{2}m,\label{eq:mCos3equalUp2}\end{equation}
By Lemma \ref{lem:m_cos3Equal_m_up2isEquivalentToOrder4} this implies
the conclusion.

Vice versa, if $\phi$ is an infinitesimal of order less than or equal
4, then by Lemma \ref{lem:m_cos3Equal_m_up2isEquivalentToOrder4}
we obtain \eqref{eq:mCos3equalUp2} and we can go over again the previous
passages in the opposite direction to prove \eqref{enu:waveEqInTheorem}.

Now, let us suppose that $\phi(x,t)\in I_{4}$ for every $(x,t)\in(a,b)\times[0,+\infty)$,
then\begin{align}
\text{lenght}(\gamma_{-,t}) & =\int_{a}^{b}\sqrt{\left[1+\alpha(t)\right]^{2}+\left[\frac{\partial\gamma_{2}}{\partial x}(x,t)\right]^{2}}\diff{x}\nonumber \\
 & =\int_{a}^{b}\sqrt{1+2\alpha(t)+\left[\frac{\partial\gamma_{2}}{\partial x}(x,t)\right]^{2}}\diff{x},\label{eq:length1}\end{align}
because $\alpha(t)\in I_{2}$ and hence $\alpha(t)^{2}=0$. But $\left[1+\alpha(t)\right]\cdot\sin\phi=\frac{\partial\gamma_{2}}{\partial x}(x,t)\cdot\cos\phi$,
so\begin{align*}
\frac{\partial\gamma_{2}}{\partial x}(x,t) & =\left[1+\alpha(t)\right]\tan\phi\\
 & =\left[1+\alpha(t)\right]\left(\phi+\frac{\phi^{3}}{3}\right)\\
 & =\phi+\frac{\phi^{3}}{3}+\alpha(t)\cdot\phi,\end{align*}
because $\alpha(t)\in I_{2}$ and $\phi\in I_{4}$ and hence $\alpha(t)\cdot\phi^{3}=0$.
Substituting this in \eqref{eq:length1} and using the derivation
formula for the function $x\mapsto\sqrt{1+x}$ we obtain\begin{align*}
\sqrt{1+2\alpha(t)+\left[\frac{\partial\gamma_{2}}{\partial x}(x,t)\right]^{2}} & =1+\frac{1}{2}\cdot\left\{ 2\alpha(t)+\left[\frac{\partial\gamma_{2}}{\partial x}(x,t)\right]^{2}\right\} \\
 & =1+\alpha(t)+\frac{1}{2}\left[\phi+\frac{\phi^{3}}{3}+\alpha(t)\cdot\phi\right]^{2}\\
 & =1+\alpha(t)+\frac{\phi^{2}}{2}+\frac{\phi^{4}}{3}+\alpha(t)\cdot\phi^{2}.\end{align*}
Therefore\begin{align}
\text{length}(\gamma_{-,t}) & =\int_{a}^{b}\left[1+\alpha(t)+\frac{\phi(x,t)^{2}}{2}+\frac{\phi(x,t)^{4}}{3}+\alpha(t)\cdot\phi(x,t)^{2}\right]\diff{x}\nonumber \\
 & =b-a+\alpha(t)\cdot(b-a)+\int_{a}^{b}\left[\frac{\phi(x,t)^{2}}{2}+\frac{\phi(x,t)^{4}}{3}+\alpha(t)\cdot\phi(x,t)^{2}\right]\diff{x}.\label{eq:lastIntegralAboutLength}\end{align}
Using the Theorem \ref{thm:infPolynomialsWithSmoothCoefficients}
it is not hard to prove that the last integral in \eqref{eq:lastIntegralAboutLength}
is an infinitesimal of order less than or equal 2, so the conclusion
follows from the hypothesis $\alpha(t)\in I_{2}$.$\qedNoNewLine$

\noindent \textbf{Proof of Lemma \ref{lem:equality2ImpliesEqualityOfDifferentials}:}
First of all, from the hypothesis $f(x)=_{2}g(x)$ for every $x\in(a,b)$,
we get that \begin{equation}
\st{f(x)}=\st{g(x)}\quad\forall x\in(a,b).\label{eq:stdPartOf_f_g_areEqual}\end{equation}
Now, let us fix a point $x\in(a,b)$. From Theorem \ref{thm:infPolynomialsWithSmoothCoefficients}
we obtain that we can write\begin{align*}
f(x_{1}) & =a_{0}(x_{1})+\sum_{i}p_{i}\cdot a_{i}(x_{1})\\
g(x_{1}) & =b_{0}(x_{1})+\sum_{j}q_{j}\cdot b_{j}(x_{1}),\end{align*}
for every $x_{1}\in(x-\delta,x+\delta)\subseteq(a,b)$ and where $p_{i}$,$q_{j}\in D_{\infty}$
and $a_{i}$, $b_{j}$ are ordinary smooth functions defined in an
open neighbourhood $V$ of $\st{x}\in(a,b)\cap\R$. From \eqref{eq:stdPartOf_f_g_areEqual}
we have $a_{0}(\st{x_{1}})=b_{0}(\st{x_{1}})$ for every $x_{1}\in\ext{V}$
so that $a_{0}=b_{0}$ on $V$ and hence also $\ext{a_{0}}=\ext{b_{0}}$
on $\ext{V}$. Therefore\begin{equation}
f(r)-g(r)=\sum_{i}p_{i}\cdot a_{i}(r)-\sum_{j}q_{j}\cdot b_{j}(r)\quad\forall r\in(a,b)\cap\R.\label{eq:fMinusg}\end{equation}
This difference must have order less than or equal 2 because $f(r)=_{2}g(r)$,
so\[
\omega\left[\sum_{i}p_{i}\cdot a_{i}(r)-\sum_{j}q_{j}\cdot b_{j}(r)\right]=\max_{i}\omega\left[p_{i}\cdot a_{i}(r)\right]\vee\max_{j}\omega\left[q_{j}\cdot b_{j}(r)\right]\le2.\]
Let us suppose, for simplicity, that $\omega(p_{1}\cdot a_{1}(r))$
is this term of maximum order. Because $a_{1}(r)\in\R$ it must be
that $\omega(p_{1})\le2$ and hence also $\omega(p_{i})\le\omega(p_{1})\le2$
and $\omega(q_{j})\le\omega(p_{1})\le2$. Finally we have\begin{align*}
f(x+h)-f(x) & =h\cdot f'(x)\\
 & =h\cdot a'_{0}(x)+\sum_{i}h\cdot p_{i}\cdot a'_{i}(x),\end{align*}
but $a'_{0}(x)=b'_{0}(x)$ because $a_{0}=b_{0}$ and $h\cdot p_{i}=0$
because $\omega(h)<2$ and $\omega(p_{i})\le2$; we hence obtain\begin{align*}
f(x+h)-f(x) & =h\cdot b'_{0}(x)\\
 & =h\cdot b'_{0}(x)+\sum h\cdot q_{j}\cdot b'_{j}(x)\\
 & =h\cdot g'(x)\\
 & =g(x+h)-g(x).\end{align*}
$\qedWithFinalEq$

\noindent \textbf{Proof of Lemma \ref{lem:m_cos3Equal_m_up2isEquivalentToOrder4}:}
If $m\cdot\cos^{3}h=_{2}m$, then the standard parts of both sides
must be equal\textbf{ }\[
\st{\left(m\cdot\cos^{3}h\right)}=\st{m}\]
\[
\st{m}\cdot\cos^{3}\left(\st{\phi}\right)=\st{m}.\]
By hypotheses $m$ is invertible, hence $\st{m}\ne0$ and we obtain
that $\st{h}=0$ because $0\le h\le\pi$, i.e. $h\in D_{\infty}$.
Moreover, from infinitesimal Taylor's formula applied to $\cos h$,
and from $m\cdot\cos^{3}h=_{2}m$ we obtain\begin{align*}
m\cdot\left(1-\sum_{1\le i<\frac{\omega(h)+1}{2}}(-1)^{i}\frac{h^{2i}}{(2i)!}\right)^{3} & =_{2}m\\
m\cdot\left(1+a\cdot h^{2}\right)^{3} & =_{2}m\\
m\cdot\left(1+a^{3}h^{6}+3ah^{2}+3a^{2}h^{2}\right) & =_{2}m\\
m\cdot\left(1+\alpha\cdot h^{2}\right) & =_{2}m\end{align*}
where $a:=-\sum_{1\le i<\frac{\omega(h)+1}{2}}(-1)^{i}\frac{h^{2i-2}}{(2i)!}\in\ER$
and $\alpha:=3a^{2}+3a+a^{3}h^{4}$ are invertible Fermat reals. From
this we get $m\cdot\alpha\cdot h^{2}=_{2}0$ and hence $h^{2}=_{2}0$,
i.e. $\omega(h^{2})\le2$ and hence $\omega(h)\le4$.

Vice versa, if $h$ is an infinitesimal of order less than or equal
4 (so that $\phi^{n}=0$ if $n\ge5$) we have\begin{align*}
\cos^{3}h & =\left(1-\frac{h^{2}}{2}+\frac{h^{4}}{4!}\right)^{3}=\\
 & =1-3\frac{h^{2}}{2}+3\frac{h^{4}}{4!}.\end{align*}
Therefore, $m\cdot\cos^{3}h=m-3mh^{2}\cdot\left(\frac{1}{2}-3\frac{h^{2}}{4!}\right)$
so that $m\cdot\cos^{3}h-m=-3mh^{2}\cdot\left(\frac{1}{2}-3\frac{h^{2}}{4!}\right)$
is an infinitesimal of order $\omega(h^{2})\le2$, i.e. $m\cos^{3}h=_{2}m$.$\qedNoNewLine$

The reader with a certain knowledge of SDG had surely noted that this
deduction of the wave equation cannot be reproduced in SDG because
of the use of non standard smooth functions, of the use of equalities
up to $k$-th order infinitesimals and because of the frequent use
of the useful Theorem \ref{thm:productOfPowers} to decide products
of powers of nilpotent infinitesimals.

\section{Conclusions}

The problem to turn informal infinitesimal methods into a rigorous
theory has been faced by several authors. The most used theories,
i.e. NSA and SDG, require a good knowledge of Mathematical Logic and
a strong formal control. Some others, like Weil functors (see e.g.
\citet{Kr-Mi2}) or the Levi-Civita field (see e.g. \citet{Sha})
are mainly based on formal/algebraic methods and sometimes lack the
intuitive meaning. In this initial work, we have shown that it is
possible to bypass the inconsistency of SIA with classical logic modifying
the Kock-Lawvere axiom (see e.g. \citet{Lav}) and keeping always
a very good intuitive meaning. We have seen how to define the algebraic
operations between this type of nilpotent infinitesimals, infinitesimal
Taylor formula and order properties. In the final part we have seen
several elementary examples of the use of these infinitesimals, some
of them taken from classical deductions of elementary Physics. In
our opinion, these examples are able to show that some results that
frequently may appear as unnatural in a standard context, using Fermat
reals can be discovered, even by suitably designed algorithm. Moreover,
our generalization of the classical proof of the wave equation have
shown that a rigorous theory of infinitesimals permits to obtain results
that are not accessible using only an intuitive approach.

\clearpage{}\bibliographystyle{plainnat}
\bibliography{IWL-bib}

\noindent 
\end{document}